\documentclass{WileyMSP-template}

\usepackage[colorlinks=true,citecolor=blue,urlcolor=blue,linkcolor=blue,bookmarksopen]{hyperref}
\usepackage{siunitx}
\usepackage{chemformula} 
\usepackage{graphicx}
\usepackage{dcolumn}
\usepackage{bm}
\usepackage{amsmath}
\usepackage{amsfonts}
\usepackage{mathtools}
\usepackage{color}
\usepackage{soul}
\usepackage{gensymb}
\usepackage{ulem}
\usepackage[utf8]{inputenc}
\usepackage{textcomp}
\DeclareUnicodeCharacter{2212}{\textminus}

\usepackage{setspace}
\justifying

\begin{document}

\pagestyle{fancy}
\rhead{\includegraphics[width=2.5cm]{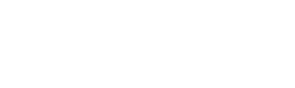}}


\title{Highly Crystalline CsPbBr$_3$ Thin Films with Amplified Spontaneous Emission via High-Pressure Recrystallization}

\maketitle


\author{Asma Miled\textsuperscript{§}}

\author{Trong Tam Nguyen\textsuperscript{§}}

\author{{\color{black}Thi Anh Vui Nguyen}}

\author{Jos\'e Penuelas}

\author{Aziz Benamrouche}

\author{C\'eline Chevalier}

\author{Thi Kim Anh Hoang}

\author{Gaëlle Trippé-Allard}

\author{Elsa Cassette}

\author{Brice Devif}

\author{Emmanuel Drouard}

\author{Emmanuelle Deleporte}

\author{Hong Hanh Mai}

\author{Abdelaziz Bouazizi}

\author{Christian Seassal}

\author{Hai Son Nguyen*}


\dedication{{§}These authors contributed equally to this work}

\begin{affiliations}

    Dr. Hai Son Nguyen\\
    Ecole Centrale de Lyon, INSA Lyon, Universit\'e  Claude Bernard Lyon 1, CPE Lyon, CNRS, INL, UMR5270, Ecully 69130, France\\
    CNRS-International-NTU-Thales Research Alliance (CINTRA), IRL 3288, Singapore 637553\\
    
    Email address: hai-son.nguyen@ec-lyon.fr

    Dr. Asma Miled\\
    Universit\'e Paris-Saclay, ENS Paris-Saclay, CNRS, CentraleSup\'elec, Lumi\`ere, Mati\`ere et Interfaces (LuMIn, UMR 9024), bâtiment 505, rue du Belv\'ed\`ere, 91400 Orsay, France\\ 
    Group of Organic Electronic and Molecular Photovoltaics Devices, Laboratory of Condensed Matter and Nanosciences, Faculty of Sciences of Monastir, University of Monastir, 5019, Monastir, Tunisia\\
    Ecole Centrale de Lyon, INSA Lyon, Universit\'e  Claude Bernard Lyon 1, CPE Lyon, CNRS, INL, UMR5270, Ecully 69130, France\\

    Dr. Trong Tam Nguyen, {\color{black}Thi Anh Vui Nguyen}, Dr. Jos\'e Penuelas\, Dr. Aziz Benamrouche, Dr. C\'eline Chevalier, Dr. Brice Devif, Dr. Emmanuel Drouard, Dr. Christian Seassal\\
    Ecole Centrale de Lyon, INSA Lyon, Universit\'e  Claude Bernard Lyon 1, CPE Lyon, CNRS, INL, UMR5270, Ecully 69130, France\\

    Thi Kim Anh Hoang\\
    Universit\'e Paris-Saclay, ENS Paris-Saclay, CNRS, CentraleSup\'elec, Lumi\`ere, Mati\`ere et Interfaces (LuMIn, UMR 9024), bâtiment 505, rue du Belv\'ed\`ere, 91400 Orsay, France\\
    Ecole Centrale de Lyon, INSA Lyon, Universit\'e  Claude Bernard Lyon 1, CPE Lyon, CNRS, INL, UMR5270, Ecully 69130, France\\
    Department of Quantum Optics, Faculty of Physics, VNU University of Science, 334 Nguyen Trai, Hanoi 100000, Vietnam\\

    Dr. Gaëlle Trippé-Allard, Dr. Elsa Cassette, Prof. Emmanuelle Deleporte\\
    Universit\'e Paris-Saclay, ENS Paris-Saclay, CNRS, CentraleSup\'elec, Lumi\`ere, Mati\`ere et Interfaces (LuMIn, UMR 9024), bâtiment 505, rue du Belv\'ed\`ere, 91400 Orsay, France\\

    Dr. Hong Hanh Mai\\
    Faculty of Electronics and Telecommunications, VNU University of Engineering and Technology, 144 Xuan Thuy, Hanoi 100000, Vietnam\\    
    Department of Quantum Optics, Faculty of Physics, VNU University of Science, 334 Nguyen Trai, Hanoi
    100000, Vietnam\\

    Dr. Abdelaziz Bouazizi\\
    Group of Organic Electronic and Molecular Photovoltaics Devices, Laboratory of Condensed Matter and Nanosciences, Faculty of Sciences of Monastir, University of Monastir, 5019, Monastir, Tunisia\\

\end{affiliations}



\begin{abstract}

Metal halide perovskites are promising materials for optoelectronic applications owing to their outstanding optical and electronic properties. Among them, all-inorganic perovskites such as CsPbBr$_3$ offer superior thermal and chemical stability. However, obtaining high-quality CsPbBr$_3$ thin films via solution processing remains challenging due to the low solubility of the precursor, and current additive or solvent engineering strategies are often complex and poorly reproducible. High-pressure recrystallization has recently emerged as a promising route to improve film quality, yet its impact on film properties remains insufficiently explored. Here, we systematically investigate the morphological, structural, and optical properties of CsPbBr$_3$ thin films prepared by high-pressure recrystallization, in comparison with standard non-recrystallized films. Optimized recrystallization at 300 bar produces smooth, pinhole-free, single-phase 3D perovskite layers with sub-nanometer roughness, while the film thickness is precisely tunable via precursor concentration. The process enhances both grain and crystallite sizes, leading to amplified spontaneous emission with a reduced excitation threshold and improved photostability. Temperature-dependent X-ray diffraction further reveals the orthorhombic--tetragonal--cubic phase transition, consistent with single-crystal behavior. This study provides fundamental insights into pressure-driven recrystallization and establishes a reproducible, scalable approach for fabricating high-quality CsPbBr$_3$ films for optoelectronic devices.

\end{abstract}


\section{Introduction}
	
Metal halide perovskites have emerged as leading materials for a broad range of optoelectronic applications, including solar cells \cite{Dastgeer2024}, light-emitting diodes (LEDs) \cite{Chen2024}, lasers \cite{Shi2025}, photodetectors \cite{Ou2024,Zhang2024}, and scintillators \cite{Anand2024}. Their exceptional optical and electronic properties, combined with low-temperature, solution-processable fabrication, have enabled rapid progress in device performance \cite{Diouf2023,Wu2021,Brenner2016}. Nevertheless, the long-term operational stability of hybrid organic–inorganic perovskites remains limited by the chemical fragility of organic components and halide-related degradation under moisture and heat \cite{Ma2022,Sun2021,Yao2021}. While passivation, surface engineering, and structural tuning can mitigate these effects \cite{Cao2022,MohdYusoff2021}, they often increase process complexity and do not fully remove intrinsic stability bottlenecks.

All-inorganic perovskites—particularly cesium lead bromide (CsPbBr$_3$)—have therefore attracted increasing interest due to their improved thermal and chemical stability \cite{Maziviero2024}. However, producing CsPbBr$_3$ thin films with the uniformity and structural quality required for integrated photonics and optical gain remains challenging. The low solubility of CsPbBr$_3$ precursors in common solvents frequently yields incomplete coverage, pinholes, and mixed-phase crystallization, which increase optical scattering and nonradiative losses \cite{Park2023,Zhang2019,Huang2019,Gupta2020}. Solvent/additive engineering and anti-solvent dripping can improve nucleation and film formation \cite{Wu2017,Yan2022,Yang2022}, but these strategies are often sensitive to processing conditions, difficult to reproduce across laboratories, and raise concerns regarding scalability and the use of toxic solvents \cite{Zhou2024}. Developing an additive-free and reproducible route to phase-pure, low-scattering CsPbBr$_3$ gain films is thus a key materials challenge for perovskite photonics.

Pressure-assisted processing has recently emerged as a promising alternative to solvent-based routes, with reports of improved morphology and enhanced optoelectronic performance. Pourdavoud et al. showed that applying $\sim$100 bar during annealing can promote smoother films with larger grains and improved lasing performance \cite{Pourdavoud2019}, while Tatarinov et al. demonstrated enhanced photostability under ambient conditions using pressure-assisted recrystallization \cite{Tatarinov2023}. However, existing studies have largely emphasized qualitative improvements or limited processing windows, and a quantitative understanding of how pressure governs film continuity, phase purity, crystallinity, and optical gain in a unified framework is still lacking.

In this work, high-pressure recrystallization of CsPbBr$_3$ thin films is investigated systematically using an automated imprinting platform that provides precise control of pressure and temperature. By scanning the recrystallization pressure (100–300 bar) and precursor concentration, a quantitative processing–structure–function map is established, linking surface morphology (coverage and roughness), phase purity and crystallinity (X-ray diffraction and peak broadening), and optical gain metrics (amplified spontaneous emission threshold and high-fluence photostability). Optimized recrystallization yields dense, pinhole-free films with sub-nanometer roughness and enhanced crystallinity, enabling low-threshold excitonic amplified spontaneous emission (ASE). Importantly, the improved structural quality further allows temperature-dependent X-ray diffraction to resolve bulk-like orthorhombic–tetragonal–cubic phase-transition signatures in a thin-film geometry, in marked contrast with standard non-recrystallized layers. Overall, this work establishes imprint-assisted pressure recrystallization as a scalable, additive-free, and environmentally benign route to high-quality all-inorganic perovskite gain films, providing a practical platform for next-generation perovskite photonics and optoelectronic devices.

\section{Results and Discussions}
\subsection{Pressure-engineered morphology and structural quality}
High-pressure recrystallization is used here as a controlled processing parameter to convert standard spin-coated CsPbBr$_3$ films into dense, low-scattering gain layers (see Experimental Methods). A quantitative pressure-property relationship is established by comparing non-recrystallized films with recrystallized films processed at 100-300~bar (fixed imprint temperature 150°C) for three PbBr$_2$ precursor concentrations (0.23, 0.30, and 0.40~M). 

\subsubsection{Non-recrystallized films:  incomplete coverage and mixed-phase crystallization}

Non-recrystallized CsPbBr$_3$ films exhibit discontinuous morphologies characterized by randomly distributed grains and a high density of pinholes, as shown in the representative scanning electron microscope (SEM) image in Figure~\ref{fig:morphology}a for 0.23 M; similar trends are also observed in Figure {\color{black}S3} for 0.30 and 0.40 M. Atomic force microscopy (AFM) measurements, shown in Figures~\ref{fig:morphology}b, {\color{black}S3b and S3e}, confirm a strongly corrugated surface with root mean square (RMS) roughness values of 25.82 nm (0.23 M), 30.80 nm (0.30 M), and 36.42 nm (0.40 M), measured over 4 $\mu$m $\times$ 4 $\mu$m areas. The average grain size extracted from image analysis is on the order of a few hundred nanometers (e.g., 242 $\pm$ 46 nm for 0.23 M; Figures {\color{black}{S8-S9}}), consistent with the heterogeneous nucleation and limited coalescence typical of solution-processed CsPbBr$_3$ under these conditions.

X-ray diffraction (XRD) further indicates that non-recrystallized films exhibit mixed crystalline phases. For 0.23 M, diffraction peaks associated with 3D CsPbBr$_3$ coexist with features attributed to a 0D secondary phase (Cs$_4$PbBr$_6$) (Figure~\ref{fig:morphology}c) \cite{Saidaminov2016}. This is indicated by the intense diffraction peaks at $2\theta = 15.21^\circ, 21.50^\circ, 30.60^\circ,$ and $37.7^\circ$ for the 3D phase and the weaker peaks at $2\theta = 12.6^\circ$ and $25.4^\circ$ for the 0D phase. At 0.30 M, an additional peak at $2\theta = 35.1^\circ$ consistent with a 2D CsPb$_2$Br$_5$ phase appears (Figure{\color{black}{S3c}}) \cite{Bai2023}, while at 0.40 M the film approaches a predominantly 3D CsPbBr$_3$ pattern (Figure{\color{black}{S3f}}). Overall, the non-recrystallized route yields films whose incomplete coverage and/or mixed-phase character are expected to increase scattering and nonradiative loss channels, limiting their suitability for optical gain and photonic applications.

\subsubsection{High-pressure recrystallization: pinhole suppression, sub-nanometer roughness and phase purity}

\begin{figure}[htb!]
\begin{center}
    \includegraphics[width=\linewidth]{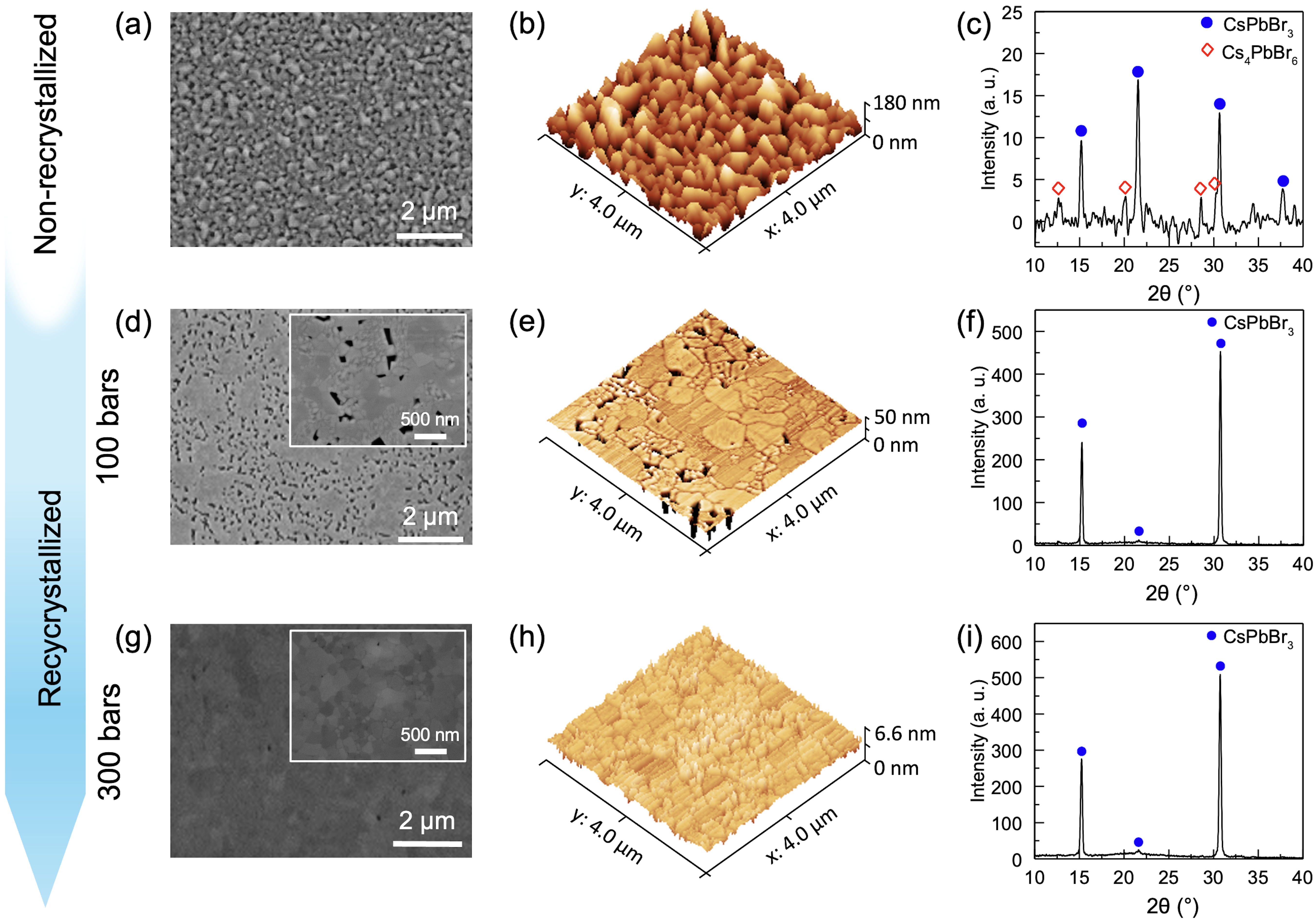}
\end{center}
	\caption{(a-c) are the SEM image, AFM image, and XRD spectrum of the non-recrystallized perovskite film, respectively. (d-f) and (g-i) are the SEM images, AFM images, and XRD spectra of the recrystallized perovskite film under 100 bar and 300 bar, respectively. The precursor concentration of PbBr$_2$ is 0.23 M.}
	\label{fig:morphology}
\end{figure}

\begin{figure}[htb!]
\begin{center}
	\includegraphics[width=0.6\linewidth]{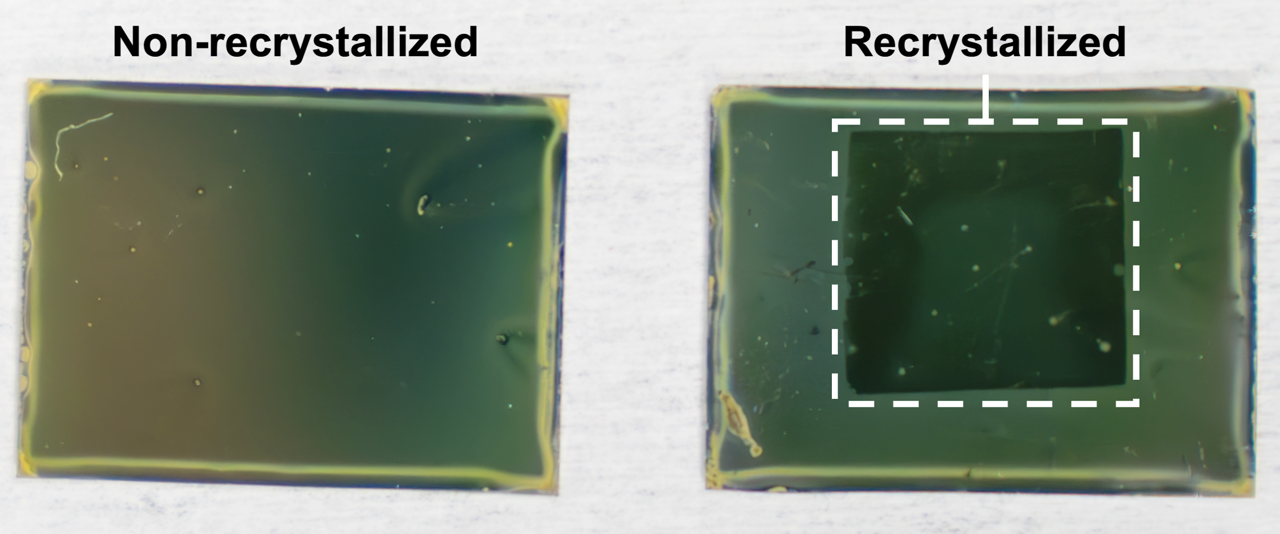}
    \end{center}
	\caption{Optical image of the non-recrystallized (left) and the recrystallized (right) CsPbBr$_3$ thin film under 300 bar of pressure, both samples correspond to the PbBr$_2$ precursor concentration of 0.23 M.}
	\label{fig:optical_image}
\end{figure}

Applying high pressure during recrystallization produces an immediate macroscopic signature: the pressed region (defined by the mold area) becomes glossy and mirror-like, while the unpressed region remains diffuse (Figure~\ref{fig:optical_image}). The results suggest that high pressure has flattened the perovskite crystals, leading to a continuous layer. Microscopically, SEM shows that recrystallization substantially increases film continuity. At 100 bar (0.23 M), pinholes are strongly reduced but still present (Figure~\ref{fig:morphology}d), whereas at 300 bar the surface becomes nearly pinhole-free (Figure~\ref{fig:morphology}g). However, it should be noted that prolonged exposure to e-beams during SEM imaging can damage perovskite films and may introduce artifacts such as pinholes. {\color{black} Notably, the higher-pressure (350 and 400 bar) recrystallized layers also exhibit similar smooth and homogeneous morphology, as shown in Figure S2, depicting the saturation of this densification. Hence, 300 bar is chosen as a sufficient recrystallizing pressure to obtain complete coverage of the CsPbBr$_3$ layer.}

    AFM measurements (Figures~\ref{fig:morphology}e and 1h) further reveal that pressure drives a dramatic reduction of surface roughness and promotes larger grains. For 0.23 M, RMS roughness decreases from 25.82 nm (non-recrystallized) to 0.67 nm after recrystallization at 300 bar (Table~\ref{tab:layer_properties}), i.e., more than one order of magnitude improvement. Similar smoothing is observed for 0.30 M and 0.40 M, yielding RMS values of 2.72 nm and 2.37 nm, respectively (Table~\ref{tab:layer_properties}; Figure {\color{black}S4b,e}). Grain sizes also increase under pressure, reaching $\sim$445 $\pm$ 69 nm at 100 bar and $\sim$511 $\pm$ 88 nm at 300 bar (0.23 M; Figure {\color{black}{S8}}). These results demonstrate that pressure-recrystallization converts rough, partially covered layers into dense, high-coverage and smooth films— essential requirements for low-loss planar waveguiding and optical gain.

    Beyond morphology, pressure-recrystallization improves crystallinity and suppresses secondary phases. XRD patterns of recrystallized films are dominated by the (hh0) peaks reflecting the characteristics of orthorhombic CsPbBr$_3$ (e.g., peaks near $2\theta \approx 15.2^\circ$ and $30.7^\circ$; Figures~\ref{fig:morphology}f and~\ref{fig:morphology}i) \cite{ref39Zhang2022}. Compared to the non-recrystallized film, the pressed film exhibits a clear narrowing of the diffraction peaks. For instance, the full width at half maximum (FWHM) of the (220) peak near $2\theta \approx 30.7^\circ$ decreases from 0.340$^\circ$ (non-recrystallized) to 0.240$^\circ$ (300 bar), indicating increased coherence length. Using the Scherrer relation $L=\frac{K\lambda}{B\cos\theta}$, where $K=0.94$ is the Scherrer constant, $\lambda=0.154$ nm is the X-ray wavelength, $B$ is the FWHM, and $\theta$ is the diffraction angle, the corresponding crystallite size increases from $\sim$28 nm to $\sim$40 nm. This enhancement confirms that high-pressure recrystallization promotes the formation of larger crystals. In parallel, the diffraction intensity increases by up to an order of magnitude, consistent with enhanced crystallinity and improved structural organization. Importantly, recrystallized films prepared at 0.30 M and 0.40 M under 300 bar are consistently single-phase within the detection limit, and show stronger peak intensities than the 0.23 M case (Figure {\color{black}S4c,f}), suggesting that higher precursor concentrations further improve crystalline quality while maintaining full coverage.

    {\color{black} Besides the impact of recrystallization pressure, annealing temperature is also reported as a factor to promote the grain growth of the perovskites \cite{Wang2022}.To better understand the effect of temperature and pressure on the recrystallization process, additional control experiments were performed. The films were treated either without pressure or under a pressure of 300 bar at different temperatures (100, 150, and 300°C), and their morphologies were subsequently compared.
    
    The SEM comparison of the non-pressed films processed at different temperatures (100°C, 150°C, and 300°C) relative to the pristine sample, as shown in Figure S5, does not reveal significant morphological changes. All samples exhibit a non-homogeneous surface  morphology with the presence of numerous pinholes, demonstrating that temperature alone is insufficient to induce complete densification of the film. In contrast, the SEM comparison (Figures S6) between pristine films and those recrystallized under 300 bar at different temperatures (100°C, 150°C, and 300°C) demonstrates that pressure plays a crucial role in promoting film densification. In particular, at 100°C, the application of high pressure significantly reduces the pinhole density, resulting in a more compact and homogeneous morphology. The increase of temperature to 150°C further enhances pressed film surface uniformity and grain growth, indicating a more advanced recrystallization stage. Interestingly, no significant morphological differences are observed between 150°C and 300°C (Figure S6d), suggesting that the system reaches a saturation point at which recrystallization is essentially complete and further increases in temperature do not notably affect grain growth. Moreover, the XRD measurements of pressed films at 150°C and 300°C exhibit similar peak position and intensity within the investigated range (Figure S7), once again supporting that the observed improvements are primarily governed by the pressure-induced recrystallization process. 
    
    As a result, the optimal recrystallization conditions are achieved at 150°C under high pressure of 300 bar, where efficient grain growth and film densification are obtained.} 
    

\subsubsection{Thickness control and pressure-induced densification}

\begin{table}[h]
\begin{center}
    \begin{tabular}{c cc cc}
        \hline
        PbBr$_2$ precursor concentration & \multicolumn{2}{c}{Non-recrystallization} & \multicolumn{2}{c }{Recrystallization (300 bar)} \\     
        (M) & Thickness (nm) & RMS (nm) & Thickness (nm) & RMS (nm) \\
        \hline
        0.23 & 120 ± 28 & 25.82 & 61 ± 9 & 0.67 \\
        0.30 & 143 ± 25 & 30.80 & 75 ± 9 & 2.72 \\
        0.40 & 163 ± 22 & 36.42 & 115 ± 8 & 2.37 \\
        \hline
    \end{tabular}
\end{center}
    \caption{Thickness and RMS values for varying PbBr$_2$ precursor concentrations. Note that RMS values are extracted from AFM over the scanned surface, whereas thickness and its standard deviation are measured from SEM at different cleaved-edge positions; therefore, the thickness deviation does not directly correspond to the RMS roughness.}  
    \label{tab:layer_properties}
\end{table}

    
   Film thickness is a key parameter for waveguiding, light absorption/trapping, and optical gain mechanisms for the performance of optoelectronic devices such as solar cells, LEDs and lasers. We measured the thickness of both the pristine spin-coated and recrystallized layers using 45°--tilted SEM images in (Figure {\color{black}{S10}}), with the results summarized in Table~\ref{tab:layer_properties}. For non-recrystallized films, thickness increases moderately with precursor concentration (120 $\pm$ 28 nm at 0.23 M to 163 $\pm$ 22 nm at 0.40 M). In contrast, recrystallized films show a stronger concentration dependence (61 $\pm$ 9 nm to 115 $\pm$ 8 nm), enabling practical thickness tuning while retaining high film quality.

    Notably, pressure-recrystallization systematically reduces thickness relative to the non-recrystallized films, consistent with densification/compaction under applied pressure. The reduction is most pronounced at low concentration (0.23 M: $\sim$120 nm $\rightarrow$ $\sim$61 nm), and less pronounced at high concentration (0.40 M: $\sim$163 nm $\rightarrow$ $\sim$115 nm), suggesting that initially more porous films exhibit stronger compaction. These results indicate that precursor concentration provides a direct handle for thickness selection, while high-pressure recrystallization simultaneously densifies and smooths out the films—an advantageous combination for photonic devices.

\subsubsection{Fiber texture in recrystallized CsPbBr$_3$ films}

\begin{figure}
\begin{center}
    \includegraphics[width=0.5\linewidth]{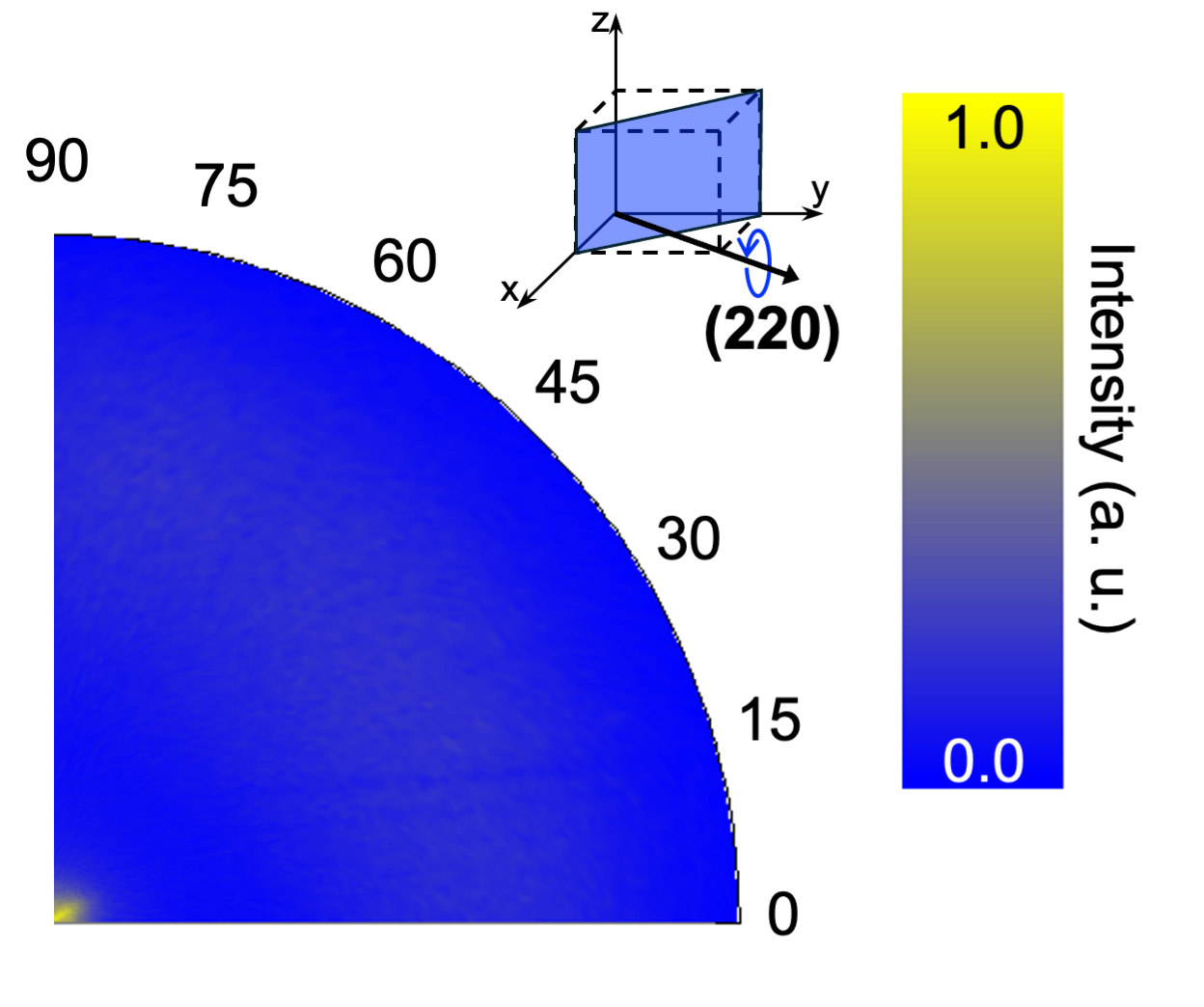}	
\end{center}
	\caption{(220) XRD pole figure of the recrystallized layer (for 0.4 M PbBr$_2$ concentration, 300 bar)}
	\label{fig:polefigure}
\end{figure}

    To assess the crystallographic texture, XRD pole-figure measurements were performed on a recrystallized CsPbBr$_3$ film (for 0.40~M PbBr$_2$ concentration, 300~bar). The (220) pole figure in Figure~\ref{fig:polefigure} displays a pronounced intensity maximum at $\chi=0^\circ$, evidencing a preferential out-of-plane alignment of crystallites. This observation is consistent with the corresponding $\theta$–2$\theta$ scans, which are dominated by (hh0) reflections, and together indicates that the recrystallized layer exhibits a fiber texture with a well-defined preferred orientation along the surface normal. The origin of this texture likely reflects the combined effect of thermodynamic driving forces (surface/interface energy minimization) and the mechanical boundary conditions imposed during pressure-assisted recrystallization, which can promote oriented grain growth under anisotropic stress. As noted by Wang et al.\cite{Wang2020}, anisotropic strain and controlled fabrication processes can play critical roles in aligning crystal orientations, particularly in perovskite thin films where quasiepitaxial growth mechanisms often emerge. Beyond enhancing structural coherence, such a texture can be advantageous for optical applications by reducing orientational disorder and potentially improving charge transport properties and light extraction efficiency, as previously reported for well-aligned perovskite layers. In this context, CsPbBr$_3$ single crystals are known to display pronounced optical anisotropy, including exciton-related birefringence effects \cite{Ermolaev2023}. While the present films remain polycrystalline, the strong preferential alignment may enable partial transfer of anisotropic optical responses at the ensemble level, which could be beneficial for devices relying on directional optical properties, such as polaritonics, lasers, and advanced light management in optoelectronic devices.

\subsection{Optical characteristics and amplified spontaneous emission}

\subsubsection{Absorption and low-fluence photoluminescence}
\begin{figure}
	\includegraphics[width=\linewidth]{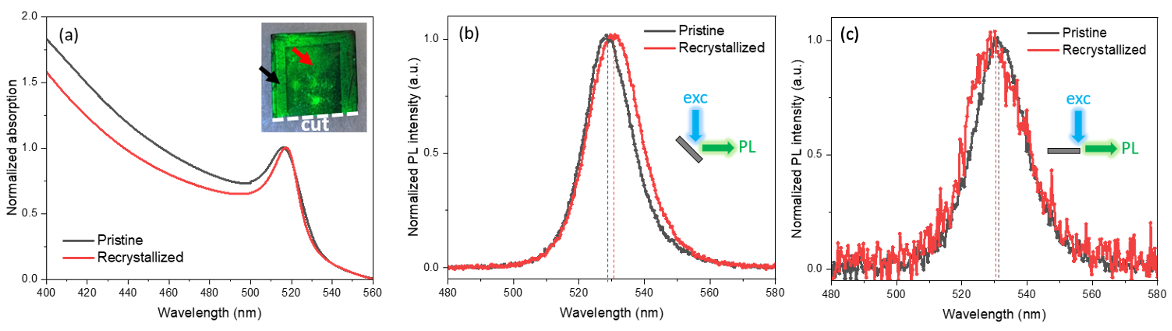}
	\caption{(a) Absorption spectrum of CsPbBr$_3$ non-recrystallized (black) and recrystallized (red) films on transparent fused silica substrate. The inset is an image of the sample on Si/SiO$_2$ substrate used for ASE measurement. (b,c) Low excitation fluence photoluminescence (PL) spectra of the non-recrystallized (black) and recrystallized (red) part of the same CsPbBr$_3$ thin film sample prepared with 0.4 M of PbBr$_2$ concentration on Si/SiO$_2$ substrate, measured in the (b) 45° and (c) 90° excitation-detection configuration, as schematized in the insets.}
	\label{fig:Absorption_PL}
\end{figure}

Understanding the optical properties of CsPbBr$_3$ thin films is crucial for optimizing their performance in optoelectronic devices \cite{Thakur2024}. Figure~\ref{fig:Absorption_PL}a shows the optical absorption spectra of the non-recrystallized and recrystallized regions of the same film prepared with 0.4M PbBr$_2$ precursor concentration. Both spectra exhibit a distinct excitonic peak, consistent with reports on single crystals \cite{Su2018} and polycrystalline films \cite{Wang2020b}. The excitonic peak is located at $\sim$516~nm for the non-recrystallized region and $\sim$517~nm for the recrystallized region, with a slight redshift and a modest narrowing after recrystallization.

Under above-bandgap photoexcitation, both regions display bright green photoluminescence (PL), as shown in the inset of Figure~\ref{fig:Absorption_PL}a and Figure~{\color{black}{S12}}. The higher apparent brightness of the non-recrystallized region arises from its rough surface, which scatters light more efficiently toward the observation direction. In contrast, the recrystallized region, characterized by a much flatter surface, promotes in-plane light confinement through total internal reflection. As a result, a larger fraction of the emitted light is guided within the film rather than radiated out of plane, yielding a comparatively dimmer appearance at this viewing angle.

To compare edge and surface emission, the sample was cleaved (white dashed line in Figure~\ref{fig:Absorption_PL}a, inset) to access both regions under 90° edge detection. Under low-fluence 400~nm excitation, the recrystallized film preserves the single-peak PL spectrum of the non-recrystallized film, as confirmed in both the 45° (Figure~\ref{fig:Absorption_PL}b) and 90° (Figure~\ref{fig:Absorption_PL}c) excitation--detection configurations. At 45° detection, the recrystallized film shows a slight redshift of the PL peak (531~nm) and a modest broadening (FWHM = 19~nm) compared with the pristine film (529~nm, FWHM = 17~nm), consistent with crystallite growth that relaxes microstrain and slightly reduces the bandgap \cite{Ummadisingu2021}. In the 90° geometry, the pristine film exhibits a $\sim$2~nm redshift relative to 45°. This additional redshift is attributed to photon recycling during in-plane propagation: in the rough, scattering-dominated pristine layer, multiple elastic-scattering events increase the effective optical path length, thereby enhancing self-absorption within the PL-absorption overlap and spectrally filtering the emission toward longer wavelengths. In contrast, the recrystallized film forms a smooth, dense planar layer that supports low-loss in-plane waveguiding by total internal reflection. The strongly reduced roughness lowers out-of-plane scattering and improves modal confinement, so that a larger fraction of the emission is funneled into guided modes rather than being diffusely scattered. At the same time, the reduced disorder (and smaller thickness) limits reabsorption during propagation, such that the edge-emitted PL remains essentially unchanged compared with the 45° collection. These waveguiding properties are consistent with the macroscopic mirror-like appearance of the pressed region and are expected to be beneficial for optical-gain operation, where reduced scattering loss and efficient in-plane propagation promote ASE build-up. Overall, the PL features are consistent with excitonic emission in polycrystalline CsPbBr$_3$ films, with a slight blueshift relative to single crystals \cite{Su2018}.

   \subsubsection{Amplified spontaneous emission: threshold and high-fluence photostability}

\begin{figure}
    \begin{center}
     \includegraphics[width=\linewidth]{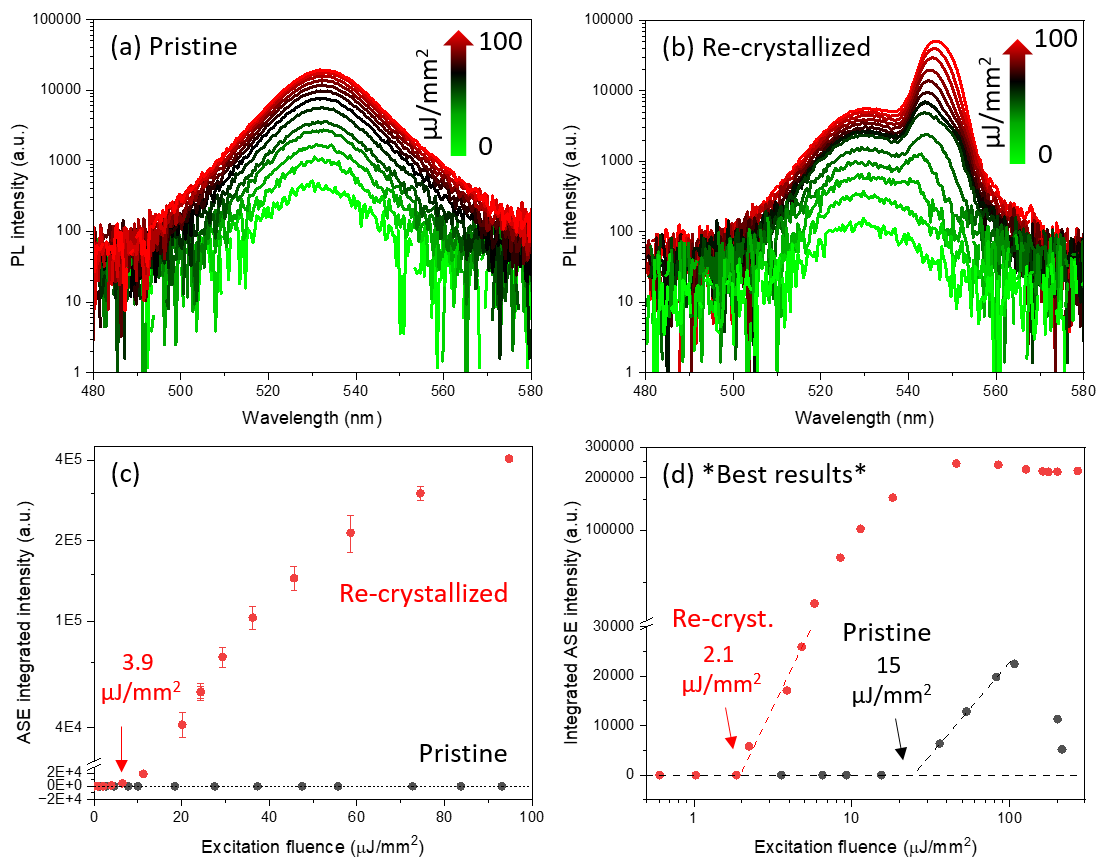}   
    \end{center}
	\caption{{PL spectra of (a) the non-recrystallized part and (b) the recrystallized part (0.4 M PbBr$_2$ concentration) with increasing excitation fluence. (c) Integrated ASE intensity plotted versus the excitation fluence, for the corresponding samples. (d) Integrated ASE intensity obtained for the best samples.}}
	\label{fig:ASE}
\end{figure}

    Under high excitation fluence (up to 200 µJ/mm$^2$,{\color{black} of a fs-laser pulse}, the recrystallized area exhibits distinct amplified spontaneous emission (ASE) behavior, which is hardly observed in the non-recrystallized part. Indeed, Figure~\ref{fig:ASE}a demonstrates typical PL spectra evolution of the non-recrystallized part according to increasing excitation fluence (up to 100 µJ/mm$^2$), showing an identical profile of the spontaneous emission with a PL peak around 529 nm. However, apart from the typical spontaneous emission peaking at 530 nm, a second sharp peak (546 nm) appears in the PL spectra of the recrystallized layer with fluence above 5 µJ/mm$^2$, as shown in Figure~\ref{fig:ASE}b, which is characteristic of ASE. The only rare non-recrystallized area found to exhibit ASE behavior (Figure {\color{black}{S13a}})is at the edge of the substrate, which is specifically higher in concentration as a result of the spin-coating process. The ASE peak of this non-recrystallized part is observed at around 539 nm at a fluence of 20 µJ/mm$^2$. Notably, all ASE measurements are performed with 90° configuration to maximize the contribution of the ASE relative to the spontaneous emission\cite{Milanese2024}. 

    To quantify ASE, the PL spectra are fitted with two Gaussian peaks, one of which corresponds to the spontaneous emission at a shorter wavelength and the other corresponds to the ASE contribution at a longer wavelength, as illustrated in Figures {\color{black}{S13c}} and {\color{black}{S13d}} in the SI. In general, the difference in energy between the center of the spontaneous emission and the position of the rising ASE is about 40 meV, which is attributed to the inter-exciton interactions \cite{Milanese2024}. Further redshift in ASE is observed at higher fluence, which can be attributed to the negative contribution of reabsorption in the optical gain.
    
    The ASE threshold is defined as the intersection of linear fits to the integrated emission intensity below and above the ASE onset.  Figure~\ref{fig:ASE}c shows a clear threshold at 3.9~$\mu$J/mm$^2$ for the recrystallized CsPbBr$_3$ layer, whereas no comparable threshold is observed for the pristine region of the same sample. Considering measurements at multiple positions, Figure~\ref{fig:ASE}d indicates a best threshold of  $\sim$15~$\mu$J/mm$^2$ for the non-recrystallized region, more than seven times higher than the best value achieved in the recrystallized region (2~$\mu$J/mm$^2$). This reduction is attributed to the enhanced crystallinity and the smoother morphology of the recrystallized layer, which reduce scattering and defect-related nonradiative losses at grain boundaries, thereby increasing the effective optical gain within the excitation volume. The larger thickness of the pristine region compared with the recrystallized region can also influence the threshold, in addition to differences in density and defect/scattering losses. The threshold values {\color{black} lie within the range commonly reported for solution-processed CsPbBr$_3$ thin films, such as CsPbBr$_3$ films doped with ammonium cations~\cite{OME_CsPbBr3_ASE_2020}, CsPbBr$_3$ nanocrystal thin films ~\cite{Balena2018,DeGiorgi2019}, CsPbBr$_3$ colloidal nanocrystals~\cite{Yakunin2015}, and bulk CsPbBr$_3$ single crystals~\cite{Zhao2019,Kim2021,Su2023}. However, the ASE threshold remains higher than other systems with improved crystalline quality or different optical geometries (film thickness and excitation geometry), including high-centimeter-scale CsPbBr$_3$ single-crystal thin films (ASE thresholds of $\sim$55.6$\mu$J/cm$^2$)~\cite{APL_CsPbBr3_SCTF_2024}), imprint-assisted recrystallized CsPbBr$_3$ layers (best threshold at $\sim$12.5$\mu$J/cm$^2$)~\cite{Pourdavoud2019}, thermally evaporated CsPbBr$_3$:TPPO (ASE threshold of 6~$\mu$J/cm$^2$)~\cite{Huang2023} (listed in table S1).}

    {\color{black} Moreover, the 0.4 M CsPbBr$_3$ thin film recrystallized at 300 bars exhibits ASE behavior under a quasi-continuous wave excitation (nanosecond pulsed laser), as shown in Figure S14a and S14b. Due to the longer pulse duration, hence, lower peak energy, the ASE thresholds are higher (4.9 $\pm$0.6$\mu$J/mm$^2$) than those obtained under fs-pulse excitation. Additionally, higher-pressure recrystallizations at 350 bar and 400 bar also exhibit ASE behavior (Figures S14c-f) with similar threshold values (Figure S15), suggesting that 300 bar is a sufficient value to achieve this high-quality, solution-based all-inorganic perovskite thin film.}
   
    Beyond the reduced threshold, the recrystallized layer also exhibits enhanced photostability at high excitation fluence (a few tens of µJ/mm$^2$). As shown in Figure~\ref{fig:ASE}d, the integrated PL reaches its saturation at a fluence of around 60 µJ/mm$^2$ and maintains an almost constant intensity under excitation beyond 100 µJ/mm$^2$. On the contrary, the ASE from the pristine part quickly degraded as the fluence exceeded 100 µJ/mm$^2$. This could be explained by the absence of "hot spots" in the recrystallized part that can occur in rough surfaces and the better crystallinity, hence allowing easier heat dissipation mediated by phonons for the recrystallized CsPbBr$_3$.

    Overall, the recrystallized film, characterized by superior film quality and well-defined grains, exhibits enhanced ASE process. The difference with the pristine part not only confirms the high optical quality of the recrystallized film but also underscores the critical role of processing conditions in dictating the emission properties of perovskite thin films.

\subsection{Phase Transition Dynamics in Recrystallized Thin Films}

\begin{figure}
	\includegraphics[width=\linewidth]{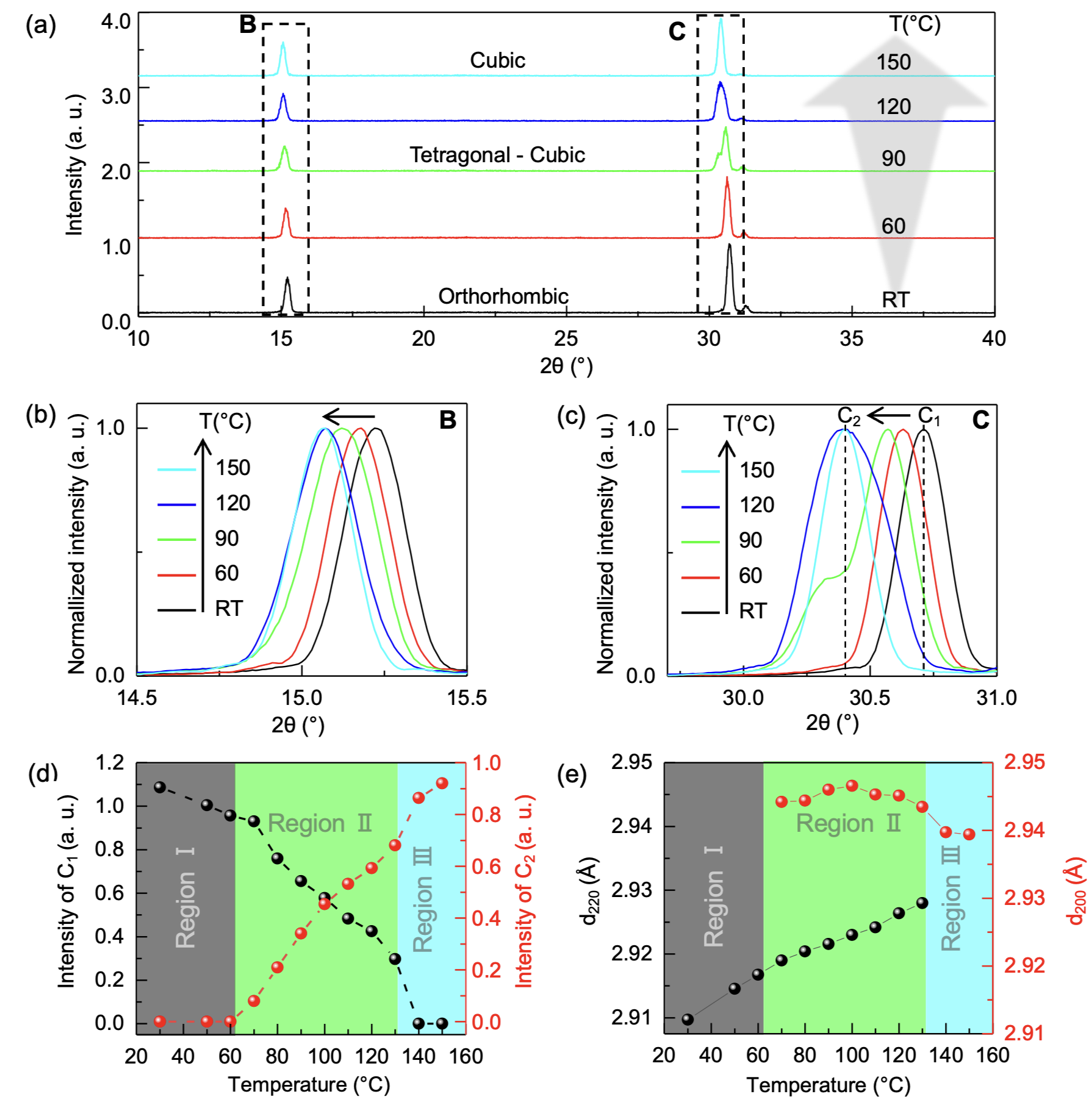}
	\caption{(a) Temperature-dependent XRD patterns of recrystallized CsPbBr$_3$ thin films prepared from the highest precursor concentration (0.4~M) at room temperature and at 60, 90, 120, and 150°C. (b,c) Enlarged, normalized views of the two principal peaks (peaks B and C) centered at $2\theta \approx 15.21^\circ$ and $30.70^\circ$, respectively. (d) Integrated intensities of the components $C_1$ and $C_2$, assigned to the (220) and (200) planes, respectively, as a function of temperature. (e) Corresponding interplanar spacings extracted from $C_1$ and $C_2$ as a function of temperature.}
	\label{fig:phase_transition}
\end{figure}

    Although structural phase transitions in CsPbBr$_3$ have been extensively reported in single crystals\,\cite{Hirotsu1974,Sharma1991,Svirskas2020,Malyshkin2020,Chen2019}, they are more rarely resolved in polycrystalline thin films\,\cite{Tenailleau2019,Whitcher2019} owing to limited crystallinity, texture broadening, and the frequent coexistence of secondary phases. Given the high crystallinity and phase purity achieved after high-pressure recrystallization, temperature-dependent XRD was performed to probe whether bulk-like orthorhombic--tetragonal--cubic signatures can be resolved in the thin-film geometry. Such an analysis is generally challenging for non-recrystallized layers in the present system, which exhibit reduced crystallinity and may contain 0D/2D secondary phases.

    Figure~\ref{fig:phase_transition}a shows XRD patterns of the recrystallized layer under 300 bars with a PbBr$_2$ precursor concentration of 0.4 M at room temperature, 60°C, 90°C, 120°C, and 150°C. The data reveal a clear shift of the Bragg peaks (B) at $2\theta = 15.21^\circ$ and (C) at $2\theta = 30.70^\circ$ to lower angles as the temperature increases. These shifts, better emphasized in Figures~\ref{fig:phase_transition}b,c, suggest a structural transformation from the orthorhombic to the cubic phase. Specifically, at T $<$ 60°C, peak B at $15.21^\circ$ corresponds to the (110) plane, and peak C at $30.70^\circ$ corresponds to the (220) plane, consistent with the orthorhombic phase. As the temperature increases to T = 150°C, peak B shifts to $15.11^\circ$ and peak C shifts to $30.40^\circ$, corresponding to the (100) and (200) planes of the cubic phase. A similar phenomenon was previously reported for epitaxially grown CsPbBr$_3$ thin films by Yifan et al\cite{Wang2019}.

    A closer inspection of peak C (Figure~\ref{fig:phase_transition}c) reveals the emergence of a shoulder around 90°C, suggesting the progressive appearance of an additional structural component during the transition. This phase transition is further characterized by examining the intensity evolution and interplanar distances of the corresponding diffraction planes (hkl), as shown in Figures~\ref{fig:phase_transition}d,e. Based on the 2$\theta$ positions, we indexed the diffraction peaks as follows: (220) for the orthorhombic and tetragonal phases (peak C$_1$) and (200) for the cubic phase (peak C$_2$). The analysis reveals three distinct regions of interest:

    \begin{enumerate}
        \item \textbf{Region I (T $<$ 60°C):} A linear evolution of the interplanar distance of the diffraction peak at $2\theta \approx 30.70^\circ$, corresponding to (220) of the orthorhombic phase.

        \item \textbf{Region II (60°C $<$ T $<$ 140°C):} A coexistence of the tetragonal and cubic phases, marked by the evolution of interplanar distances and intensities of diffraction peaks at $2\theta \approx 30.50^\circ$ (tetragonal, (220)) and $2\theta \approx 30.40^\circ$ (cubic, (200)). A significant shift in interplanar distances during this region highlights the phase transition from the orthorhombic phase to the coexistence of tetragonal and cubic phases.

        \item \textbf{Region III (T $>$ 140°C):} Suppression of the orthorhombic peak and stabilization of the interplanar distance corresponding to (200), marking the complete transition to the cubic phase.
    \end{enumerate}

    To further support our findings, we evaluated the thermal expansion coefficient $\alpha$ by fitting the linear evolution of the interplanar distance $d_{220}$ in Figure~\ref{fig:phase_transition}e. The analysis reveals two distinct values of $\alpha$, corresponding to Region I and Region II: $\alpha_I = 0.8096 \times 10^{-4}$ K$^{-1}$ and $\alpha_{II} = 0.5008 \times 10^{-4}$ K$^{-1}$. These values are significantly higher than those reported for single-crystal CsPbBr$_3$ \cite{Haeger2020}, likely due to the differences in structure and thermal behavior between recrystallized thin films and single crystals.

    Similar phase transition behavior was also observed for recrystallized layers with lower precursor concentrations of 0.23 M and 0.3 M, as shown in Supporting Information (Figure {\color{black}{S16}}). On average, the cubic phase (region III) remains as the temperature is higher than $\approx$130°C, which is in good agreement with the literature \cite{Hirotsu1974,Tenailleau2019,Sharma1991,Svirskas2020, Malyshkin2020}. On the other hand, the transition point between the orthorhombic and tetragonal phases is around 70°C, which is slightly lower than the reported value \cite{Hirotsu1974,Sharma1991,Tenailleau2019,Svirskas2020,Malyshkin2020}. Interestingly, the recrystallized films reverted to the orthorhombic structure upon cooling to room temperature, demonstrating the reversibility of the phase transition. 
    
    {\color{black} Therefore, the temperature-dependent XRD analysis highlights the enhanced crystallinity and improved phase stability induced by the high-pressure recrystallization process, which is consistent with the improved optical performance reported in this work.}

\section{Conclusion}

In conclusion, imprint-assisted high-pressure recrystallization is demonstrated as a robust, additive-free strategy to convert standard spin-coated CsPbBr$_3$ layers into homogeneous, highly crystalline thin films. By systematically varying pressure and precursor concentration, a quantitative processing--structure--function map is established. Under optimized conditions, the films become dense and fully covered, exhibit a single 3D CsPbBr$_3$ phase within the XRD detection limit, and reach sub-nanometer surface roughness, while the thickness remains readily tunable through precursor concentration. These improvements yield low-scattering planar waveguiding layers that support ASE with a strongly reduced threshold and enhanced photostability compared with non-recrystallized films. Moreover, the improved structural quality enables temperature-dependent XRD to resolve a reversible orthorhombic--tetragonal--cubic phase transition in a thin-film geometry, approaching the bulk-like behavior typically observed in single crystals. This scalable materials platform is therefore well suited for device integration across multiple domains, including solar cells, light-emitting diodes, photodetectors/sensors, optically pumped gain media and lasers, as well as exciton--polariton devices operating in the strong light--matter coupling regime.

A key outlook is that the imprint step can be made functional: replacing the flat mold with a patterned mold would enable recrystallization and nanostructuring in a single step,~\cite{Wang2017,Mayer2022,HaMyDang2024}. Such ``recrystallization-by-imprint'' could directly encode optical functionalities into CsPbBr$_3$ films---for example, subwavelength resonant patterns for perovskite metasurfaces---while preserving (and potentially enhancing) phase purity and crystalline coherence through pressure-driven reorganization. {\color{black} Additionally, this technique can be complementary to other high-quality thin films, such as ligand-assisted thermal co-evaporation CsPbBr$_3$, to further enhance emission performance~\cite{Huang2023_lasing}}. This approach could provide a practical route toward engineered dispersion and radiation channels, including directional emission~\cite{MermetLyaudoz2023,Wang2023}, polarization and chiral responses~\cite{Long2022,MendozaCarreo2023,Dang2022}, tailored waveguiding/outcoupling~\cite{Dang2024,Wang2024bis}, spatially structured optical gain~\cite{Pourdavoud2019,MermetLyaudoz2023,Kurahashi2024}, and enhancing nonlinear effects~\cite{Fan2021}, opening opportunities for compact beam shaping and new lasing and on-chip propagation/feedback schemes in all-inorganic perovskite photonics.

\section{Experimental Methods}
	
    \threesubsection{Sample Preparation} Perovskite solutions were prepared by mixing different precursor concentrations of CsBr and PbBr$_2$ in the same molar ratio (1.5:1) in dimethylsulfoxide  (DMSO). In particular, three precursor PbBr$_2$ concentrations have been studied: 0.23 M, 0.3 M and 0.4 M. The prepared solutions were stirred at 60°C in a nitrogen-filled glovebox overnight. The solutions were then filtered using a 0.2 µm PTFE filter before perovskite film deposition. In parallel, the Si/SiO$_2$ substrates were ultrasonically cleaned in acetone, ethanol, and isopropanol for 15 minutes in each solvent and subsequently blown dry with N$_2$ gas. The substrates were treated with UV-ozone for 15 minutes to remove residual organic molecules and to improve hydrophilicity. Finally, the substrates were loaded into an N$_2$-filled glovebox for the deposition of perovskite. 

	
	The pristine CsPbBr$_3$ layer is fabricated with a one-step spin coating followed by annealing, as demonstrated in Figures S1(a-b). In particular, the precursor solution was spin-coated onto a Si/SiO$_2$ substrate at 3000 rpm for 80 seconds. Subsequently, the samples were annealed at 100°C for 15 minutes to evaporate the solvent, allowing the perovskite to crystallize into a nanocrystalline layer, also referred to as a non-recrystallized layer.

    Recrystallization was performed using an imprint system (NPS300 – Smart Equipment Technology Corporation, France), which simultaneously applied high pressure to the crystallized layer using a flat silicon mold while maintaining a temperature of 150°C, as depicted in Figure S1(c). The temperature was held constant for 10 minutes while the pressure was gradually released as the system returned to room temperature. After removing the silicon mold, the perovskite layer was flattened, resulting in larger crystal sizes and improved surface coverage, as shown in Figure S1(d). This recrystallization process lasted a total of 35 minutes, bringing the entire fabrication process to under one hour. Furthermore, the area of the recrystallized layer depends solely on the size of the silicon mold, demonstrating that this method enables simple and fast fabrication of perovskite layers over a large area. In this study, we focused on examining the impact of pressure on the recrystallized layer. To ensure consistency, the silicon mold size was fixed at 1 cm × 1 cm with a thickness of 0.5 mm.
 
    \threesubsection{Characterization Tools} The structural properties of the perovskite layers were analyzed using XRD measurements performed with a Rigaku Smartlab diffractometer. This instrument is equipped with a 9 kW rotating anode and a two-reflection Ge (400) crystal monochromator, which selects the Cu $K_{\alpha1}$ radiation $(\lambda=1.54056\mathring{A})$.  

    For optical characterization, the absorption spectra of the layers were measured with a CARY UV-VIS-NIR spectrometer. A home-built macro-photoluminescence (macro-PL) setup was used to study photoluminescence (PL) and amplified spontaneous emission (ASE). We used an amplified TiSa laser (800 nm, 4 kHz, 80 fs, $>$ 1.5 mJ) as the excitation source, which is further frequency-doubled to 400 nm and focused on the sample with a focal length of 50 or 100 mm (corresponding beam section: 3.4$\times 10^{-4}$ and 1.4$\times 10^{-4}$, respectively). The PL was collected at 90° or 45° and focused on an optical fiber spectrometer using a set of lenses. For the 90° configuration, the excitation was set right at the edge of the sample to limit re-absorption effects. {\color{black}A home-built micro-PL setup is used to characterize the ASE under quasi-continuous-wave excitation, utilizing a 450 nm laser with a pulse duration of 5 ns and a repetition rate of 50 Hz. The ns-pulse laser is focused on the sample with a 10X microscope (NA = 0.42), which is also used to collect the emitted signal.}

\medskip
\textbf{Supporting Information} \par 
Supporting Information is available from the Wiley Online Library or from the author.

\medskip
\textbf{Acknowledgements} \par 
Technological processes have been performed in the Nanolyon technical platform, a member of the RENATECH+ national microfabrication network and the CARAT alliance. The authors thank the staff from the platform for helping and supporting all nanofabrication processes. This work is funded by the French National Research Agency (ANR) under the project POLAROID (ANR-24-CE24-7616-01). T.K.A.H. and E.C. thank the Integrative Institute of Materials of Université Paris-Saclay (ANR-11-IDEX-0003) and the ANR (ANR-2023-CE29-0003-01) for partial funding. 

\medskip

%

\bibliographystyle{MSP}
\bibliography{Ref}


\renewcommand{\thesection}{S\arabic{section}}
\renewcommand{\thefigure}{S\arabic{figure}}
\renewcommand{\thetable}{S\arabic{table}}
\renewcommand{\theequation}{S\arabic{equation}}
\setcounter{section}{0}
\setcounter{figure}{0}
\setcounter{table}{0}
\setcounter{equation}{0}


---SUPPLEMENTAL INFORMATION---\\

    \section{Sample fabrication}
    
	\begin{figure}[htb!]
		\begin{center}
            \includegraphics[width=16cm]{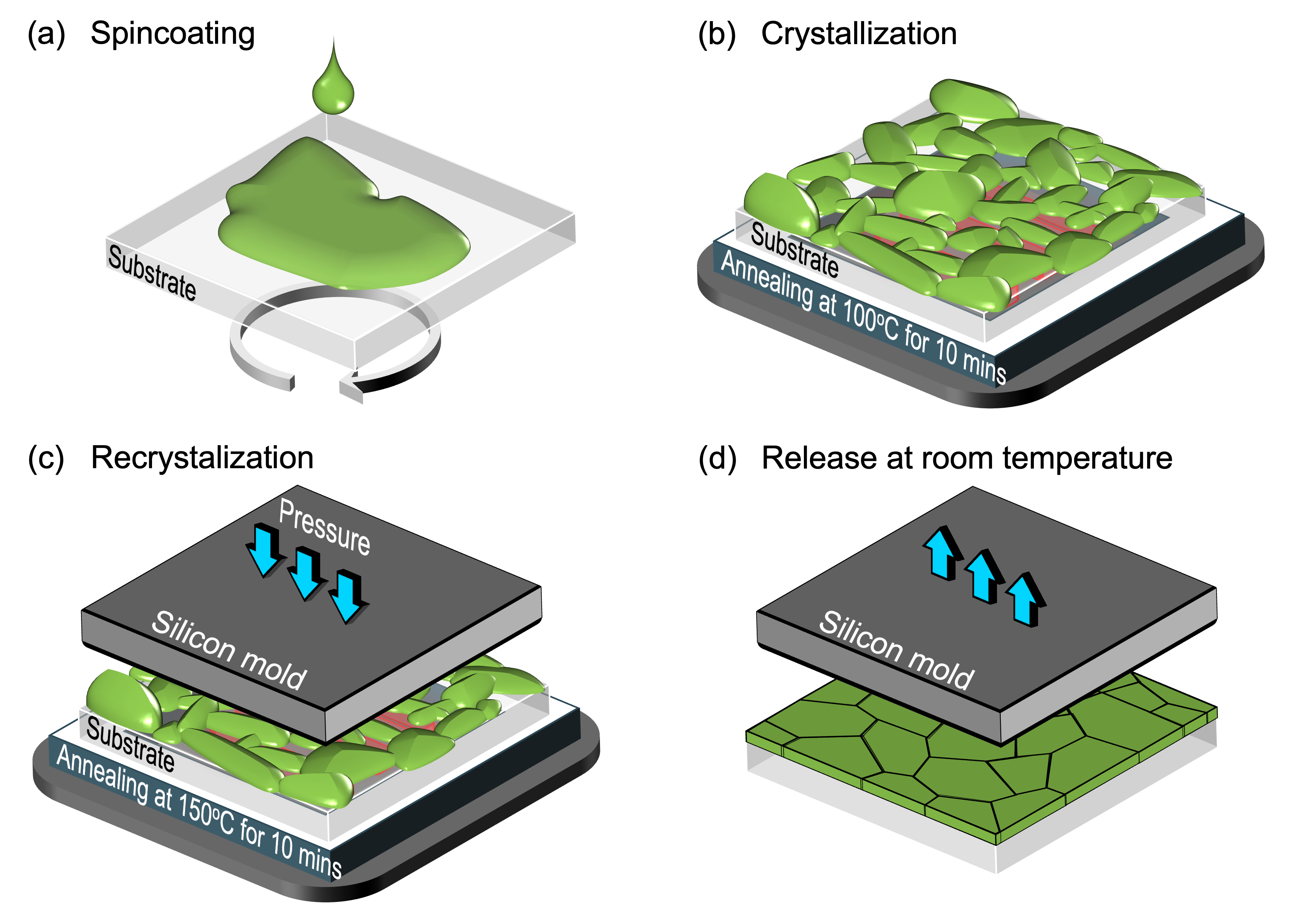}
		\end{center}
		\caption{Schematic of the two-step fabrication process: the pristine spin-coat includes (a) spin coating of the precursor solution, followed by (b) annealing to induce crystallization, also called a non-recrystallized thin film. (c) Recrystallization involves applying high pressure and a controlled crystallization temperature, resulting in (d) a flattened perovskite layer with larger crystal sizes and improved surface coverage.}
		\label{Fab}
	\end{figure}

    Figure~\ref{Fab} illustrates the fabrication process used to achieve high-quality CsPbBr$_3$ films with full coverage and high crystallinity, involving two crystallization steps. The pristine layer (Figures ~\ref{Fab}a and ~\ref{Fab}b) was first crystallized through a simple annealing process. Subsequently, the film was recrystallized (Figures~\ref{Fab}c and ~\ref{Fab}d) under both high pressure and a controlled crystallization temperature.

    {\color{black}\section{CsPbBr$_3$ layers of PbBr$_2$ precursor concentrations 0.23 M recrystallized at high pressure}}
    
    \begin{figure}[htb!]
		\begin{center}
            \includegraphics[width=15cm]{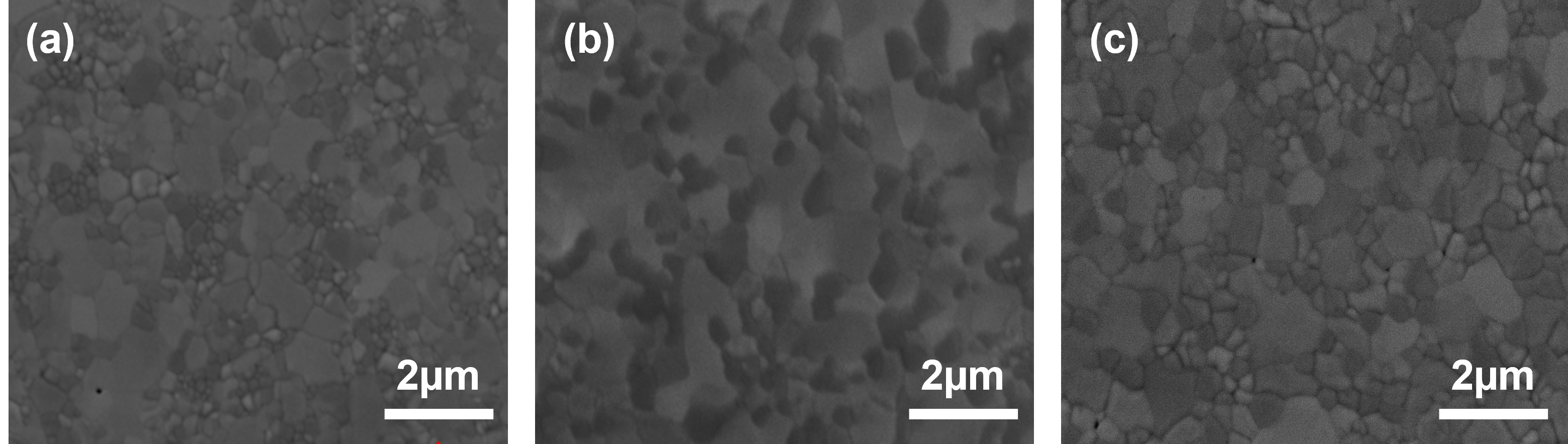}
		\end{center}
		\caption{{\color{black}SEM images of the recrystallized perovskite film under (a) 300 bar, (b) 350 bar, and (c) 400 bar.}}
		\label{SEMhigherpressures}
	\end{figure}

    {\color{black} Figure~\ref{SEMhigherpressures} shows the SEM images of the CsPbBr${_3}$ layers recrystallized under the pressure of 300, 350, and 400 bar. The film morphology at 350 and 400 bar remains essentially unchanged compared to 300 bar, suggesting a saturation of the pressure effect above 300 bar (Figure~\ref{SEMhigherpressures}).}
    
    \section{Non-recrystallized CsPbBr$_3$ layers of PbBr$_2$ precursor concentrations 0.3 M and 0.4 M}
    
    \begin{figure}[htb!]
		\begin{center}
			\includegraphics[width=16cm]{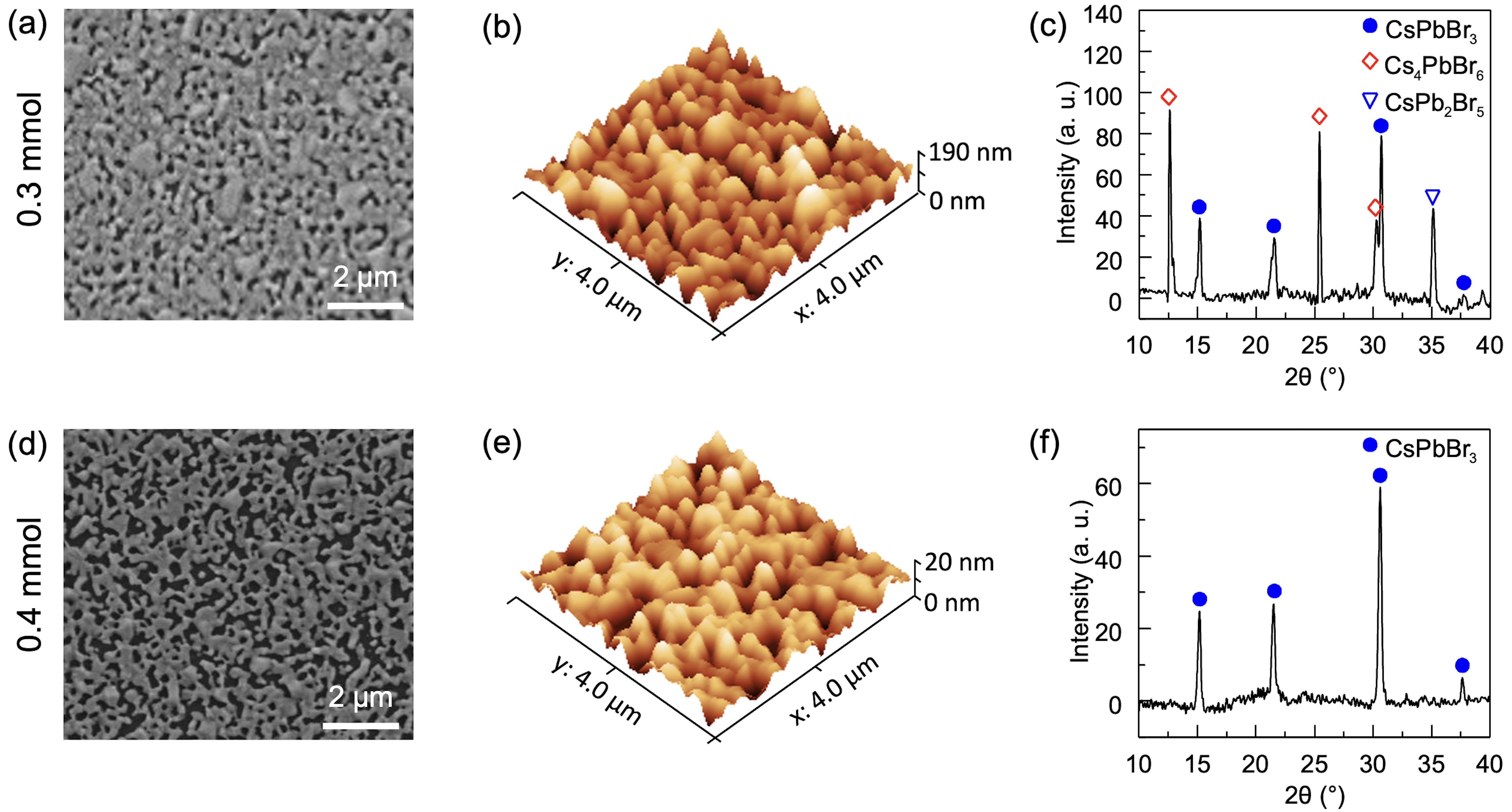}
		\end{center}
        \caption{Non-recrystallized perovskite layer of PbBr$_2$ precursor concentration 0.3 M (a) SEM image of the recrystallized layer. (b) AFM images indicate the low roughness, and (c) the corresponding XRD spectrum. Non-recrystallized perovskite layer of PbBr$_2$ precursor concentration 0.4 M. (d) SEM images of the recrystallized layer. (e) AFM images indicate the high roughness, and (f) the corresponding XRD spectrum.}
		\label{Non-Recrys-0304}
	\end{figure}

    \section{Recrystallized CsPbBr$_3$ layers of PbBr$_2$ precursor concentrations 0.3 M and 0.4 M}
    
    Figures~\ref {Non-Recrys-0304} a and ~\ref{Non-Recrys-0304}b are the SEM images and AFM measurement of the 0.3 M non-recrystallized thin film, indicating a high density of pinholes and high roughness. The XRD spectrum in Figure~\ref{Non-Recrys-0304}c shows the presence of different cesium bromide perovskite phases. This is specified by the intense diffraction peaks at 2$\theta$ = 15.21$^\circ$, 21.50$^\circ$, and 30.60$^\circ$ for the 3D phase (CsPbBr$_3$), while the peaks at 2$\theta$ = 12.6$^\circ$ and $25.4^\circ$ are for the 0D phase (Cs$_4$PbBr$_6$). The peak at $2\theta = 35.1^\circ$ reveals the presence of a 2D phase (CsPb$_2$Br$_5$) alongside the 3D and 0D phases. While exhibiting the same morphological properties as shown in Figures~\ref{Non-Recrys-0304}(d) and ~\ref{Non-Recrys-0304}e, the 0.4 M sample exhibits only a 3D phase in the thin film crystal structure (Figure~\ref{Non-Recrys-0304}f). 
    
    \begin{figure}[htb!]
		\begin{center}
			\includegraphics[width=16cm]{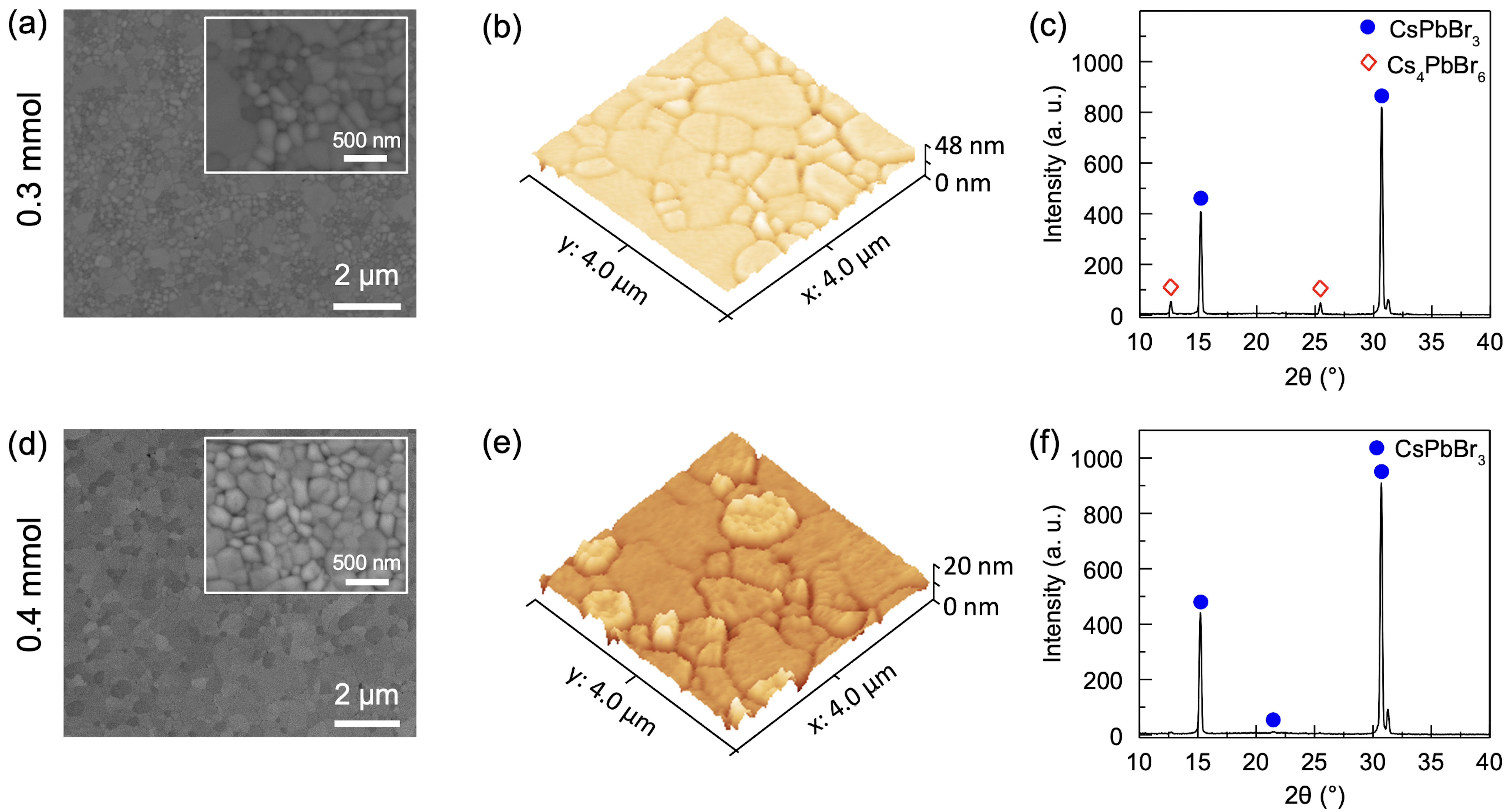}
		\end{center}
        \caption{\label{FigureS4}{Recrystallized perovskite layers under 300 bar of 0.3 M of PbBr$_2$ (a) SEM image of the recrystallized layer, the inset shows the higher-magnification SEM image of the surface. (b) AFM images indicate low roughness, and (c) the corresponding XRD spectrum. Recrystallized perovskite layer of 0.4 M of PbBr$_2$ (d) SEM images of the recrystallized layer; the inset shows the higher magnification of the surface (e) AFM images indicate low roughness and (f) the corresponding XRD spectrum.}}
		\label{Non-Recrys-0304}
	\end{figure}

    Figures~\ref {Non-Recrys-0304}a and ~\ref{Non-Recrys-0304}b, respectively, show the SEM images and AFM measurement of PbBr$_2$ precursor concentration of 0.3 M recrystallized thin film under 300 bar, which indicate pinhole-free surfaces and low roughness. The XRD spectra are shown in Figure ~\ref{Non-Recrys-0304}c, showing the high crystallinity of the recrystallized layer: the 3D phase CsPbBr$_3$ is the dominant one, while there is a minimal contribution of the 0D phase (Cs$_4$PbBr$_6$). Similar properties are observed for recrystallized thin films of PbBr$_2$ precursor concentration 0.4 M under the same pressure, as shown in ~\ref{Non-Recrys-0304}d-f. The XRD of the 0.4 M sample shows a pure 3D phase in the thin film crystal structure. These results confirm that complete coverage and high crystallinity can all be achieved with the recrystallization pressure of 300 bar. 
     
     {\color{black}\section {Temperature control of the recrystallized CsPbBr$_3$ layers of PbBr$_2$ precursor concentrations 0.23 M}}

    \begin{figure}[htb!]
		\begin{center}
            \includegraphics[width=16cm]{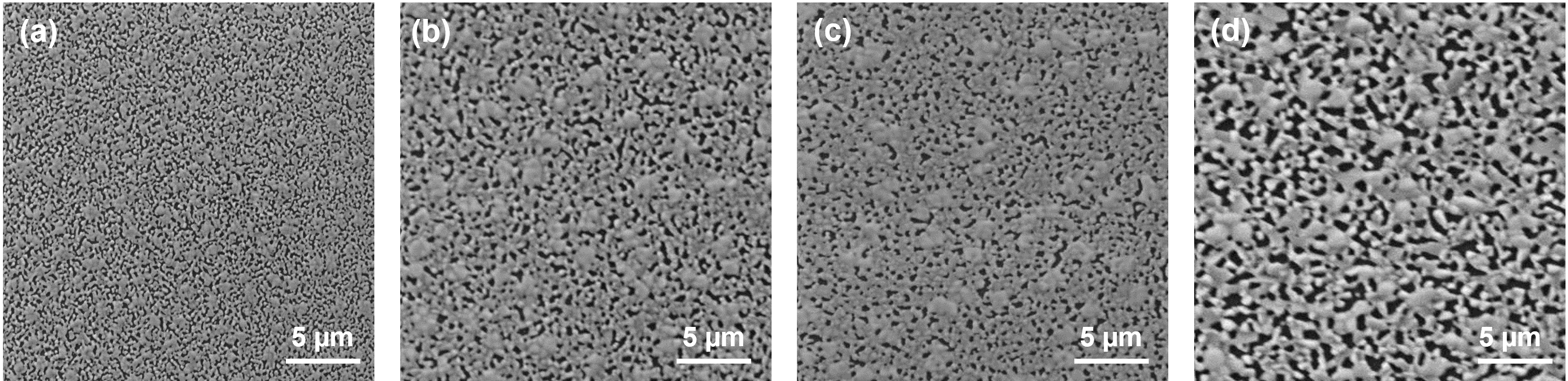}
		\end{center}
		\caption{\color{black} SEM images of (a) Pristine perovskite film and recrystallized perovskite films without pressure at (b) 100$^{\circ}$C, (c) 150$^{\circ}$C, and (d) 300$^{\circ}$C}
		\label{Recryst-Temp-Dependent}
	\end{figure}

     {\color{black} Figure~\ref{Recryst-Temp-Dependent} presents a comparative SEM analysis of 0.23 M pristine films processed at 100°C and those recrystallized films treated at different temperatures (100°C, 150°C, and 300°C)  without pressure.}

    \begin{figure}[htb!]
		\begin{center}
            \includegraphics[width=16cm]{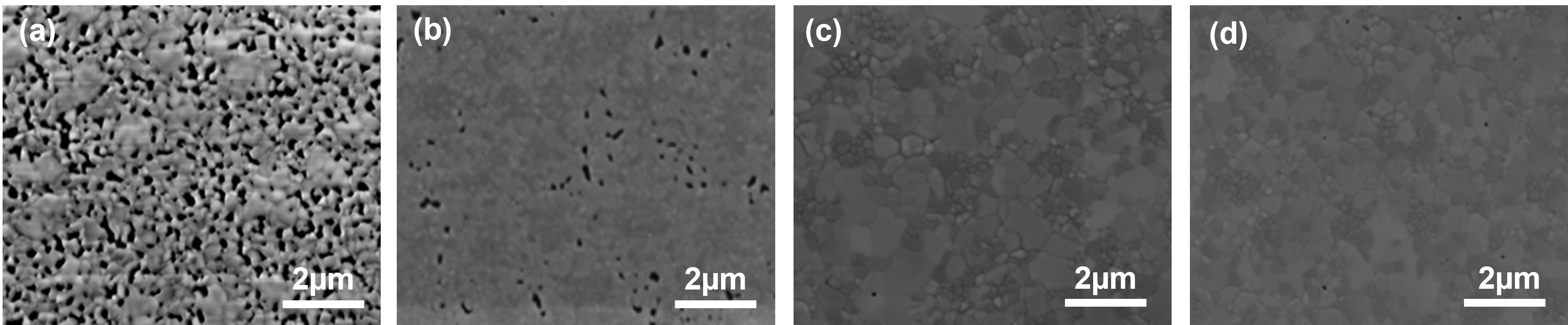}
		\end{center}
		\caption {\color{black}SEM images of (a) pristine perovskite film and recrystallized perovskite films under 300 bar at (b) 100$^{\circ}$C, (c) 150$^{\circ}$C, and (d) 300$^{\circ}$C}
		\label{Recryst-Temp-Dependent under pressure}
	\end{figure}
    
     {\color{black} Figure~\ref{Recryst-Temp-Dependent under pressure} presents a comparative SEM analysis of 0.23 M pristine film and those treated at different temperatures (100°C, 150°C, and 300°C) under an applied pressure of 300 bar.}
    
   \begin{figure}[htb!]
		\begin{center}
            \includegraphics[width=6cm]{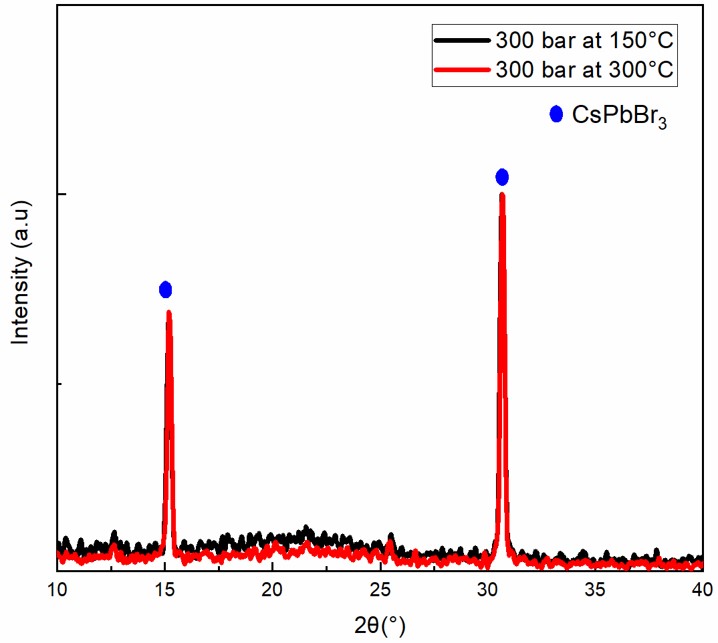}
		\end{center}
		\caption{\color{black} XRD Patterns of Recrystallized Perovskite Films under 300 bar at 150 and 300°C.} 
		\label{300bar300C_XRD}
	\end{figure}

    {\color{black}Figure~\ref{300bar300C_XRD} presents the XRD patterns of recrystallized perovskite films under 300 bar at 150°C and 300°C, highlighting the absence of structural modifications between the recrystallized layers at these two temperatures.}

    \section{Grain size distribution}
    \begin{figure}[htb!]
		\begin{center}
			\includegraphics[width=12cm]{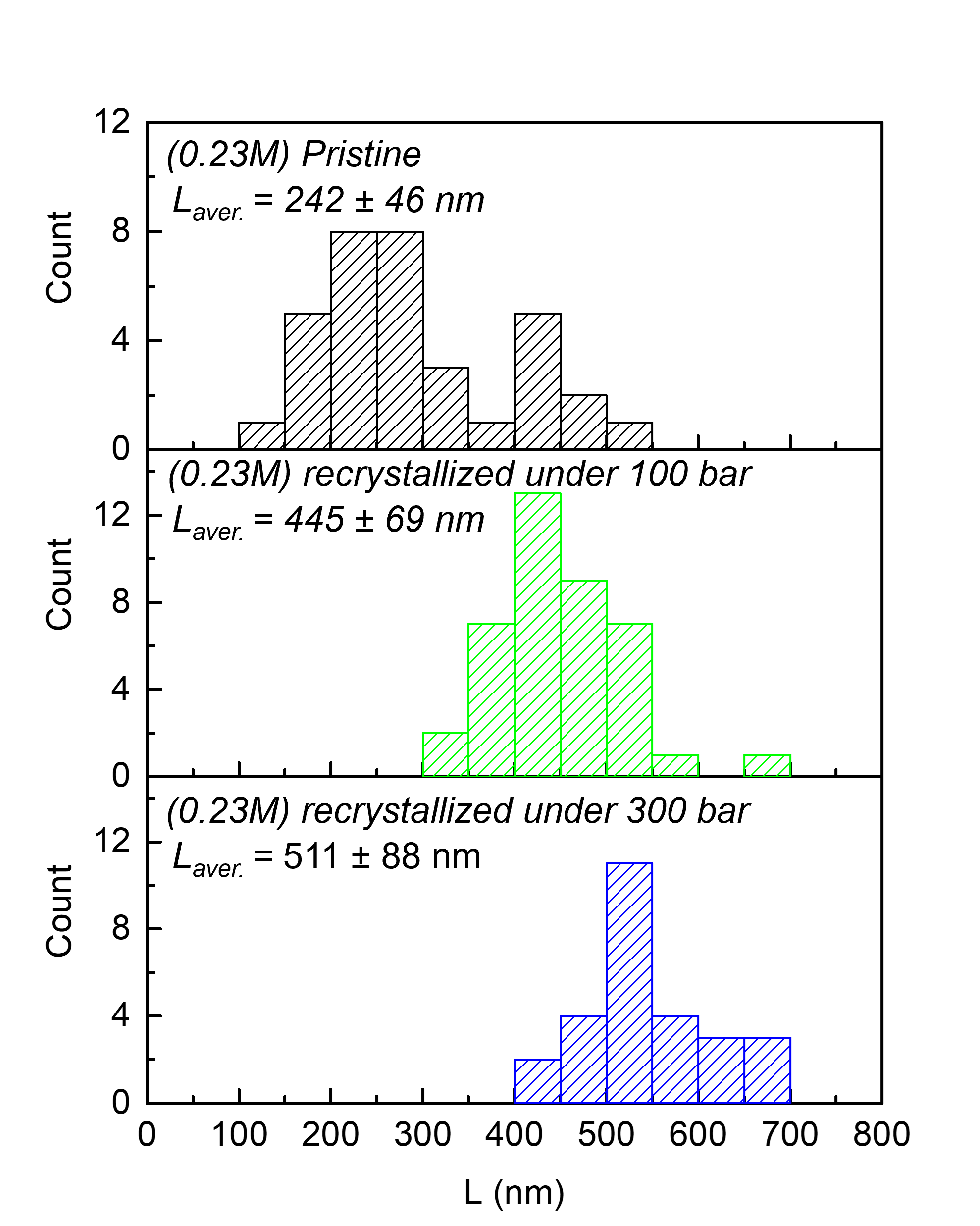}
		\end{center}
        \caption{The grain size distribution of the (top) pristine layer of 0.23 M of PbBr$_2$ concentration and of the corresponding recrystallized layer under (middle) 100 bar and (bottom) 300 bar.}
		\label{023_GrainDist}
    \end{figure}

    Figure~\ref{023_GrainDist} demonstrates the grain size distribution of the PbBr$_2$ concentration 0.23 M layers from the AFM measurements. The pristine layer exhibits an average grain size of 242 $\pm$ 46 nm (top graph). Under the high recrystallized pressure, the average grain size increases to 445 $\pm$ 69 nm at 100 bar (middle graph) and doubles to 511 $\pm$ 88 nm at 300 bar (bottom graph). The errors mentioned here refer to the standard deviation of the grain size.
    
    \begin{figure}[htb!]
		\begin{center}
			\includegraphics[width=16cm]{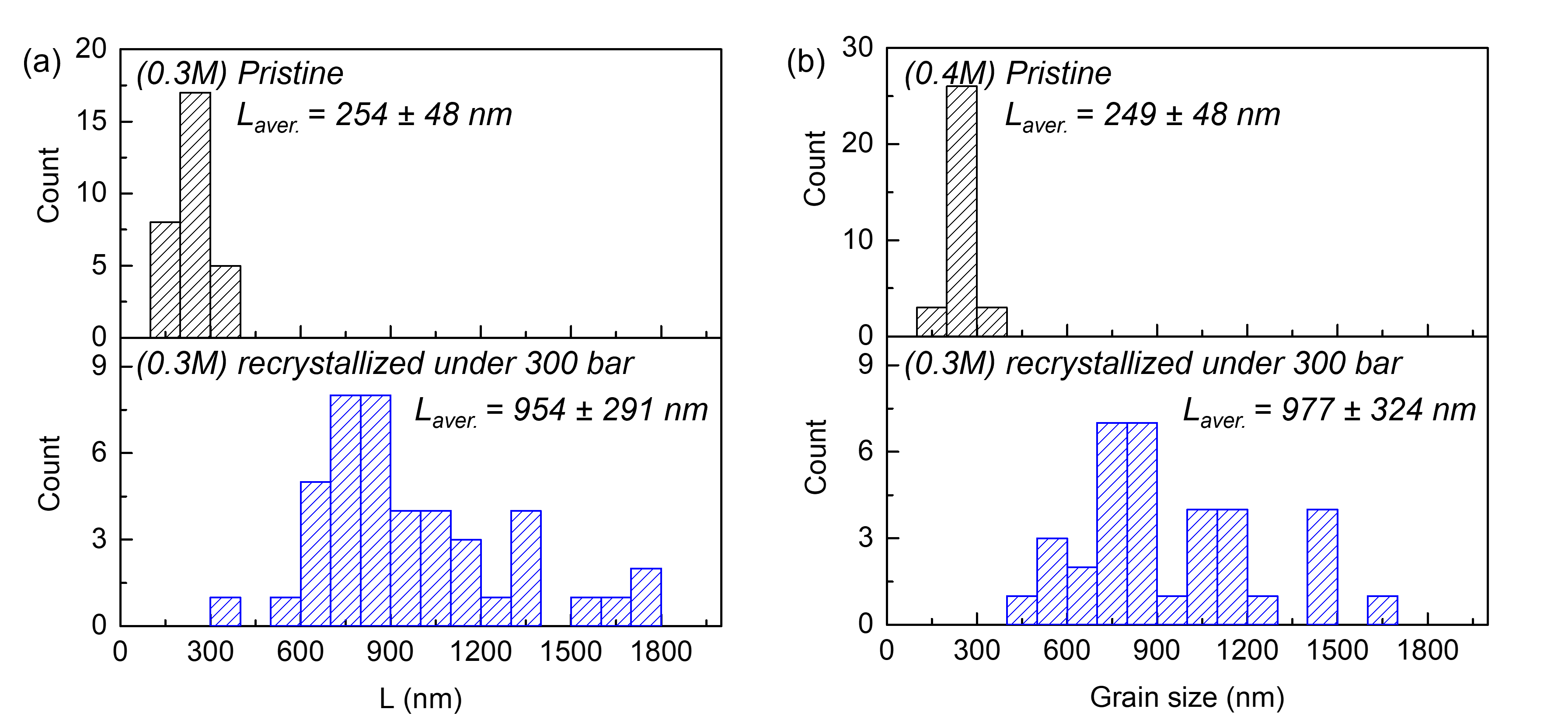}
		\end{center}
        \caption{\label{FigureS9}{(a) The grain size distribution of the (top) pristine and (bottom) recrystallized (under 300 bar) of the concentration 0.3 M of PbBr$_2$. (b) The grain size distribution of (top) pristine and (bottom) recrystallized (under 300 bar) of the concentration 0.4 M of PbBr$_2$.}}
		\label{0304_GrainDist}
    \end{figure}

    Figure~\ref{0304_GrainDist} demonstrates the grain size distribution of the 0.3 M and 0.4 M layers from their corresponding AFM measurements. Under the pressure of 300 bar, the averaged grain size of the 0.3 M layer increased from 254 $\pm$ 48 nm for the pristine layer (Figure ~\ref{0304_GrainDist}a, top graph) to 954 $\pm$ 291 nm for the pristine layer (Figure ~\ref{GrainDist}a, bottom graph). Similarly, Figure~\ref{0304_GrainDist}b shows a more than 3-fold growth of the averaged grain size of the 0.4 M layer, from 254 $\pm$48 nm for the pristine layer (top graph) to 977 $\pm$ 324 nm (bottom graph). The errors mentioned here refer to the standard deviation of the grain size.

    \section{Thickness measurements} 
\begin{figure}[htb!]
		\begin{center}
			\includegraphics[width=16cm]{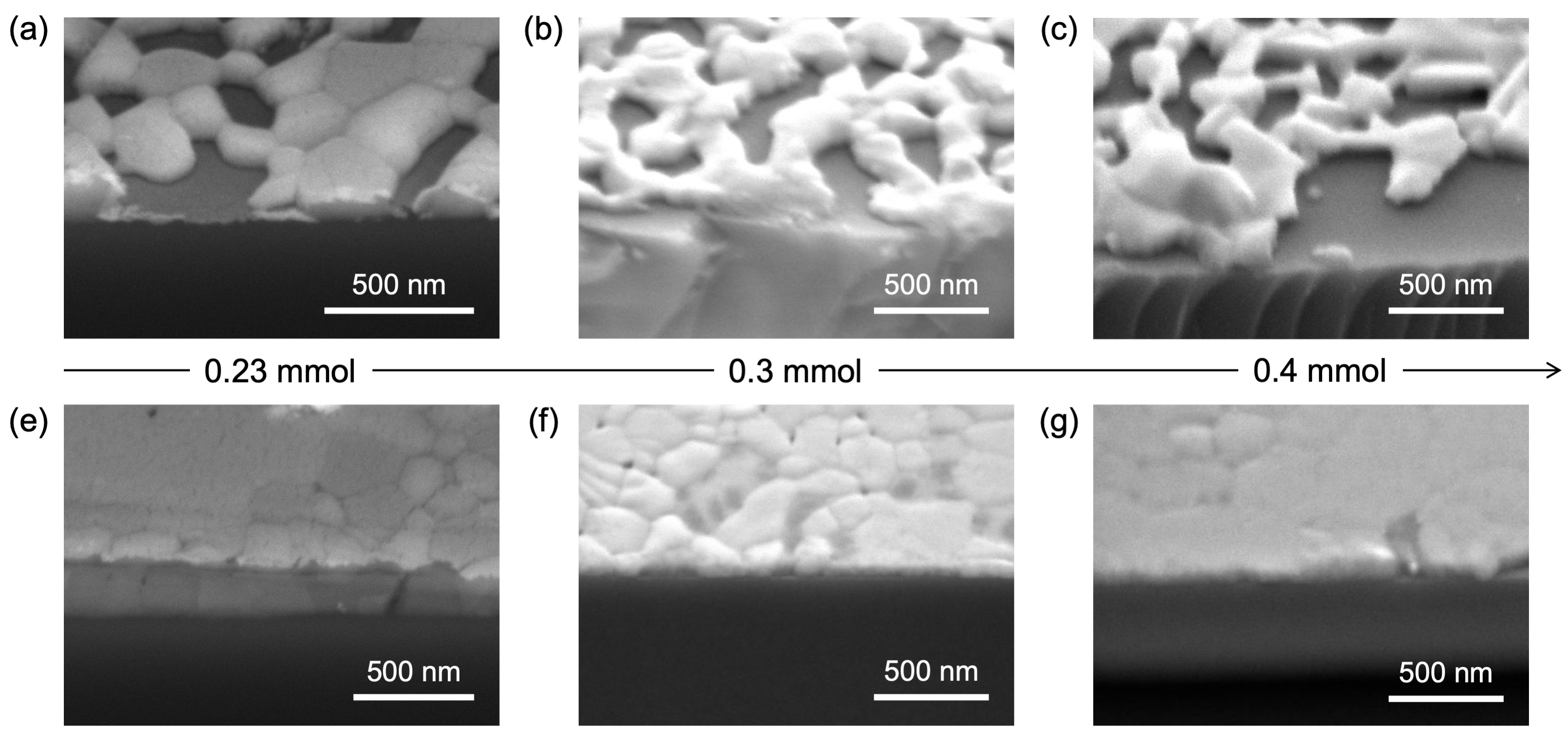}
		\end{center}
		\caption{\label{FigureS10}{(a-c) 45°- tilted SEM images of the non-recrystallized perovskite layers of the concentrations 0.23 M, 0.3 M, and 0.4 M of PbBr$_2$, respectively. (e – g) 45°- tilted SEM images of the recrystallized perovskite layers under 300 bar of concentrations of 0.23 M, 0.3 M, and 0.4 M of PbBr$_2$, respectively.}}
		\label{CrossSection}
	\end{figure}

\begin{figure}[htb!]
		\begin{center}
			\includegraphics[width=16cm]{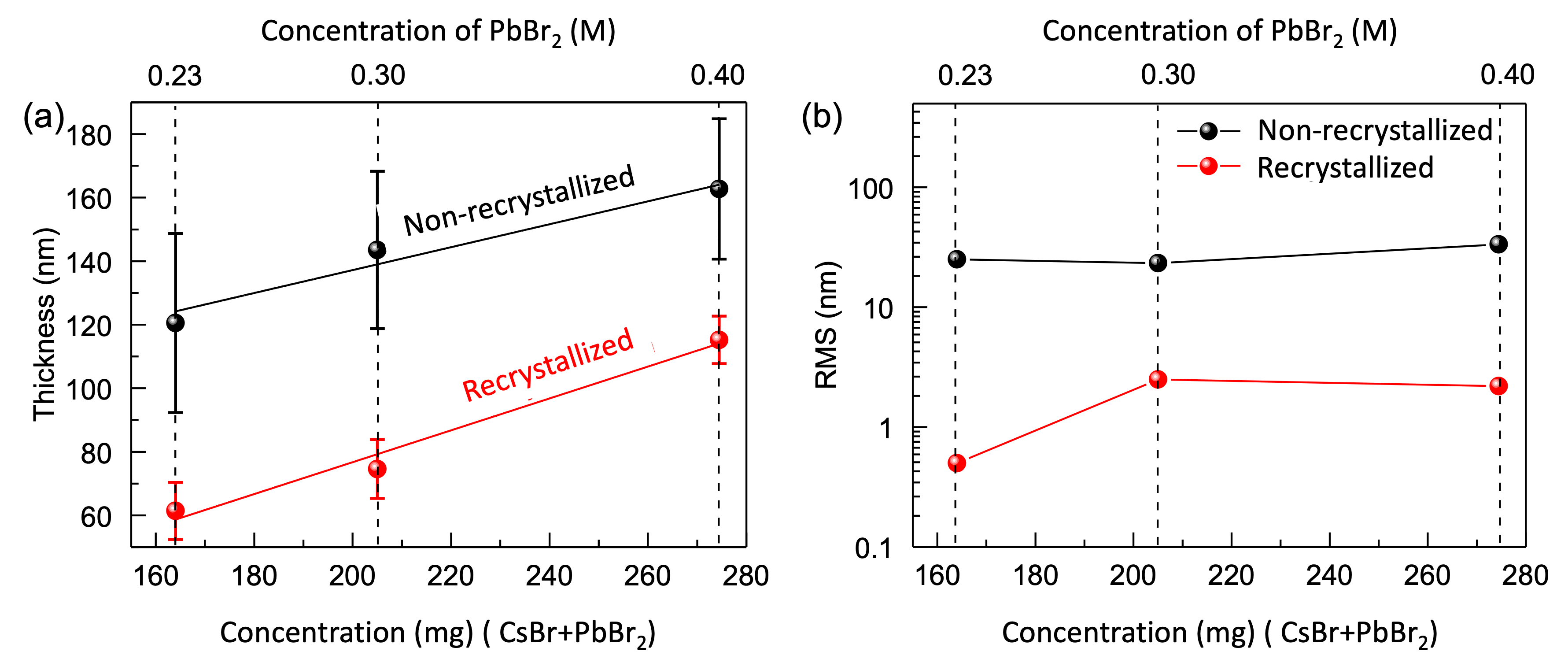}
		\end{center}
		\caption{\label{figureS11}{The evolution of (a) the thickness and (b) the roughness as a function of the PbBr$_2$ precursor and CsBr+PbBr$_2$ concentration of the non-recrystallized (black circle data) and recrystallized (red circle data) thin film}}
		\label{Roughness_Graph}
    \end{figure}

Figures~\ref{CrossSection}a-c show the 45°-tilted SEM images of the non-recrystallized thin film of PbBr$_2$ concentrations 0.23 M, 0.3 M, and 0.4 M, respectively. The same image setup of the recrystallized thin films of the PbBr$_2$ concentrations of 0.23 M, 0.3 M, and 0.4 M is also shown in Figures~\ref{CrossSection}e-g. All SEM images have the optimized magnification to avoid charging and drifting effects. The thickness of each layer is measured at the cleaved edge of the substrates at different positions. The final thickness corresponding to the precursor concentration is the average value reported in Figure~\ref{Roughness_Graph}a. 
    
    Figure~\ref{Roughness_Graph}a shows the average thickness of the non-recrystallized and recrystallized thin films as a function of PbBr$_2$ concentrations. The error bars are the standard deviation of each measurement. Under high pressure of 300 bar, the thickness of the recrystallized layer is reduced by up to 60 nm. Figure~\ref{Roughness_Graph}b shows the roughness of the non-recrystallized and recrystallized thin films as a function of the concentration of PbBr$_2$, demonstrating the significant improvement in surface roughness. 
   
    \section{Photoluminescence} 
\begin{figure}[htb!]
		\begin{center}
			\includegraphics[width=16cm]{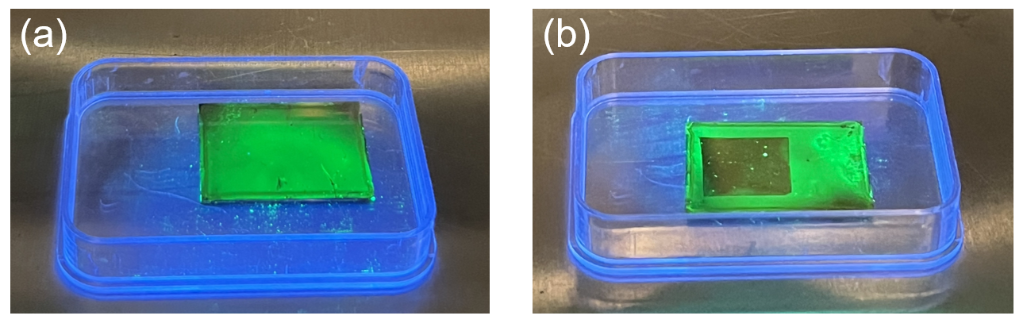}
		\end{center}
        \caption{PL images under UV-light of (a) non-recrystallized and (b) recrystallized under 300 bar CsPbBr$_3$ thin film, the corresponding PbBr$_2$ precursor concentration is 0.4 M.}
		\label{PL}
	\end{figure}

   Figures~\ref{PL}a and ~\ref{PL}b show the photoluminescence images of non-recrystallized and recrystallized (300 bar) CsPbBr$_3$ thin films under UV illumination. In Figure~\ref{PL}a, the non-recrystallized region displays a relatively uniform bright-green emission, characteristic of CsPbBr$_3$. In contrast, the recrystallized area in Figure~\ref{PL}a appears significantly dimmer than both the surrounding region and the pristine film. This apparent reduction in brightness arises from the smoother and more continuous morphology of the recrystallized layer, which favors in-plane light guiding through total internal reflection rather than out-of-plane scattering, as observed in the rougher, grainy non-recrystallized film. 
   
    \section{Amplified spontaneous emission (ASE) measurement} 
    \subsection{Under femtosecond laser excitation} 
\begin{figure}[htb!]
		\begin{center}
			\includegraphics[width=14cm]{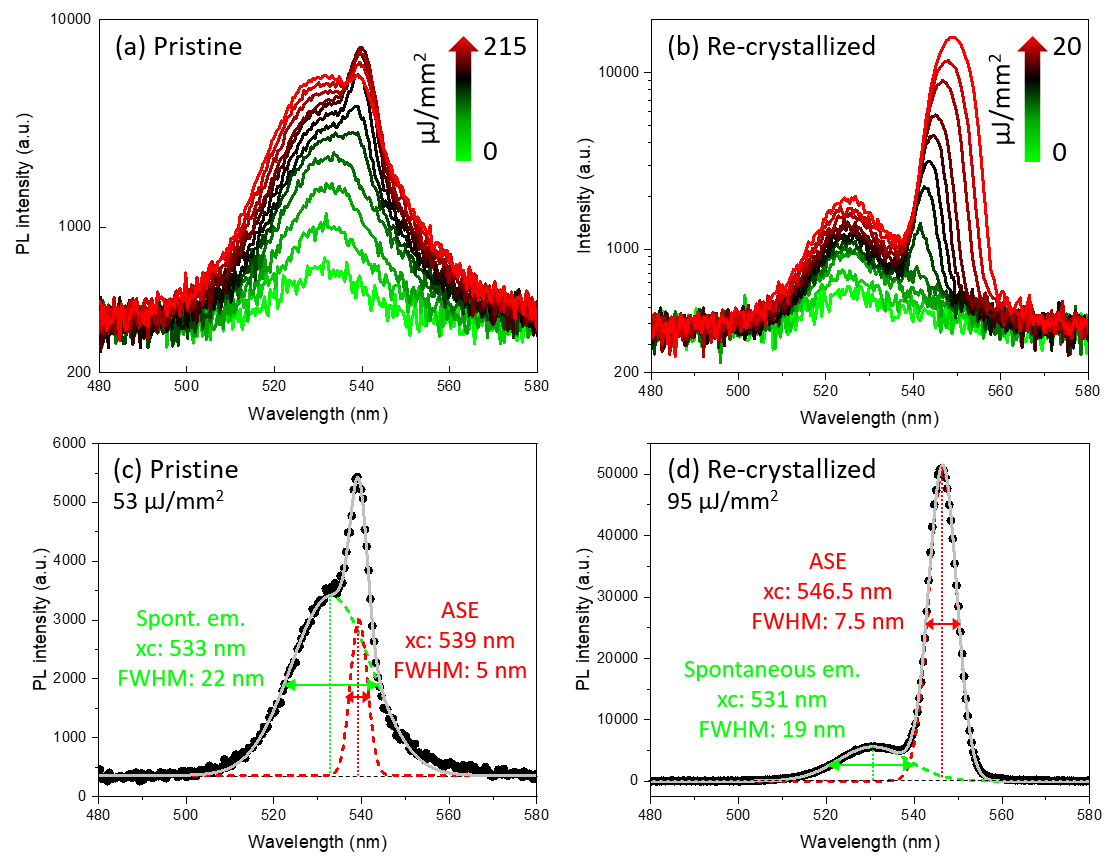}
		\end{center}
        \caption{(a) and (b) Fluence-dependent PL spectra of CsPbBr$_3$ thin films in the pristine and re-crystallized region (for 0.4 M of PbBr$_2$ precursor concentration). The spectra were recorded with a 50 mm focal length in the excitation path. (c) and (d) Examples of fits of PL spectrum at specific excitation fluences, with the spontaneous emission and ASE signals, to retrieve the center of the emission, xc, and the FWHM of the corresponding peak.}
		\label{ASEfs}
	\end{figure}

Measurements of ASE were performed in the 90° excitation-detection configuration, as shown in Figure 4c. Figures~\ref{ASEfs}a and ~\ref{ASEfs}b display the fluence-dependent PL spectra of two different samples presenting the highest ASE response. Figures~\ref{ASEfs}c and ~\ref{ASEfs}d are respectively the PL spectra of the non-recrystallized and recrystallized thin films at high excitation fluence and their corresponding two-peak fitting. In the fit, the spontaneous emission peak is located at about 533 nm for the non-recrystallized thin film and 531 nm for the recrystallized thin film, with a FWHM (Full Width at Half Maximum) of about 20 nm. The ASE part of the recrystallized thin film peaks was further red-shifted with the increase of excitation fluence and broader (from 5 to almost 8 nm). Such an effect could not be observed in the pristine part due to the rapid degradation of the film under such high excitation fluence. 

    {\color{black}\subsection{Under nanosecond laser excitation} For the ASE characterization under quasi-continuous wave excitation, the samples are optically pumped with a ns-laser source, at 450 nm. The laser is focused on the samples through a microscope objective lens, NA = 0.42, which also detects the emitted signal. To compare with previous experiments, the configuration is 0$^o$ as the excitation and detection are performed at the same angle.}
    
    \begin{figure}[htb!]
		\begin{center}
			\includegraphics[width=14cm]{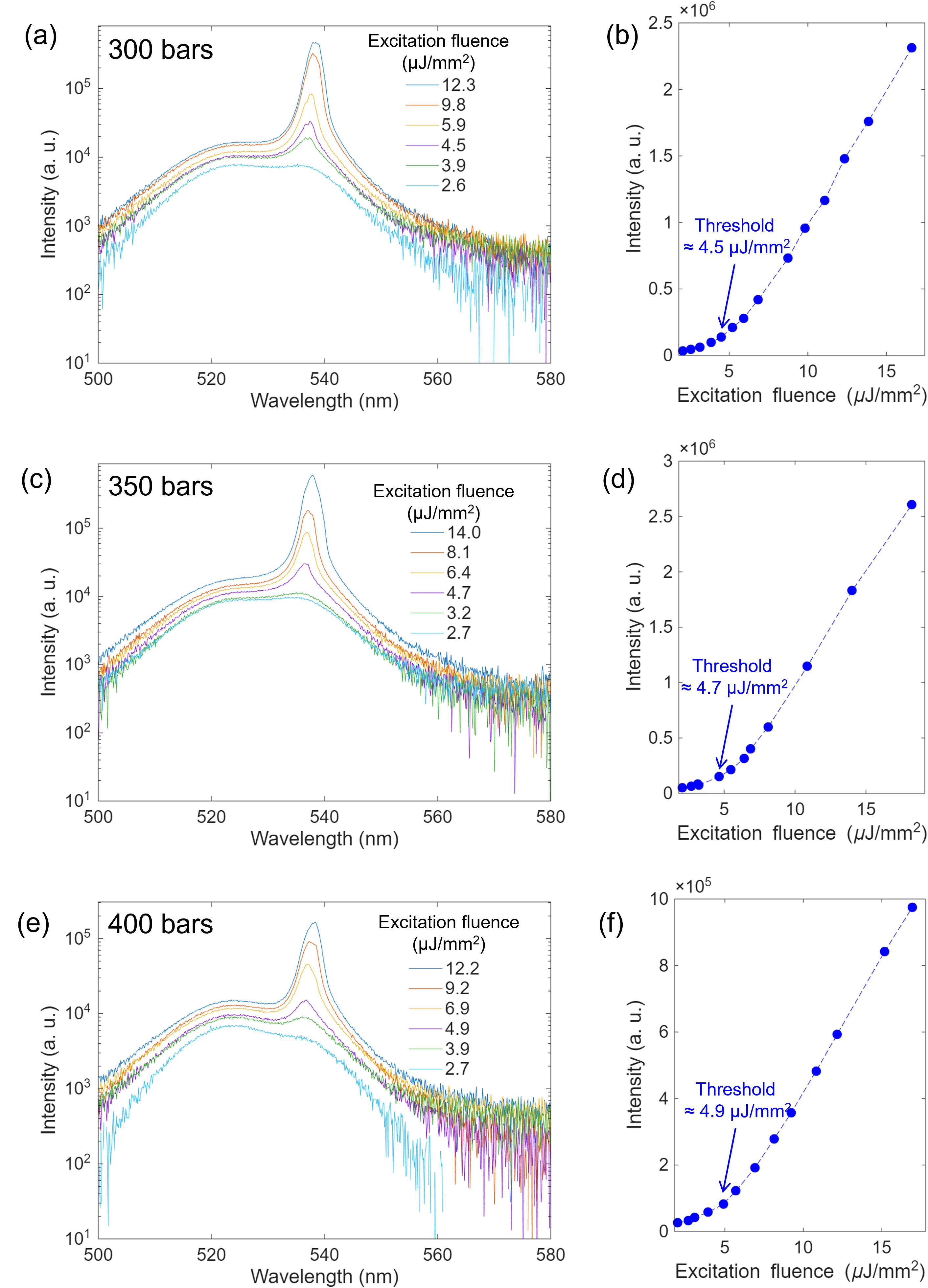}
		\end{center}
        \caption{\color{black} The fluence-dependent (a) PL spectra and (b) integrated intensity of the CsPbBr$_3$ film recrystallized at 300 bar. The fluence-dependent (c) PL spectra and (d) integrated intensity of the CsPbBr$_3$ film recrystallized at 350 bar. The fluence-dependent (e) PL spectra and (f) integrated intensity of the CsPbBr$_3$ film recrystallized at 400 bar. All experiments use an ns-pulse laser as the optical excitation source.}
		\label{ASEns}
	\end{figure}

    {\color{black}Figures~\ref{ASEns}a and ~\ref{ASEns}b demonstrate similar ASE behaviors of the CsPbBr$_3$ film recrystallized at 300 bar under increasing excitation fluence of a ns-pulse laser. Notably, due to the longer pulse duration (i.e. lower peak energy), the ASE threshold is verified around 4.5 $\mu J/mm^2$, double the value under ps-pulse excitation. Nevertheless, the clear emergence of ASE under nanosecond pumping highlights the excellent optical quality and gain robustness of the recrystallized CsPbBr$_3$ film, since sustaining population inversion over a nanosecond timescale is significantly more demanding than under ultrafast excitation.} 
    
    \begin{figure}[htb!]
		\begin{center}
			\includegraphics[width=12cm]{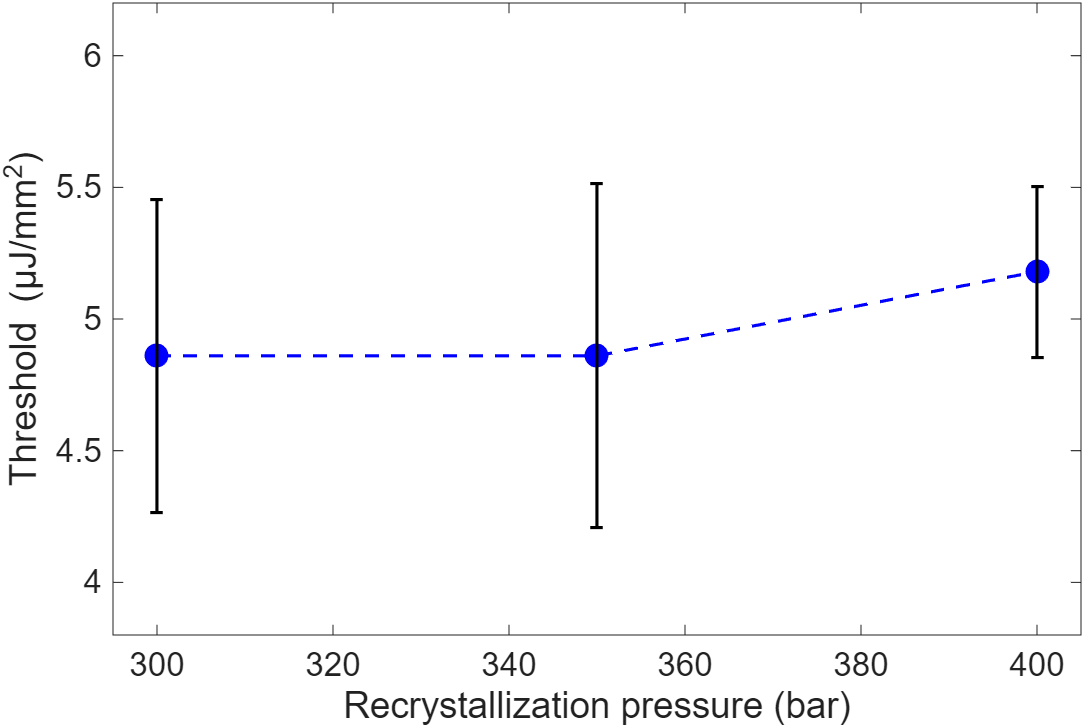}
		\end{center}
        \caption{\color{black} Average ASE thresholds of the recrystallized CsPbBr$_3$ films under 300 bar, 350 bar and 400 bar; the error bar indicate the standard deviation of the threshold measurements.}
		\label{ASEnsThres}
	\end{figure}

    {\color{black} Similarly, the CsPbBr$_3$ films recrystallized at higher pressures, 350 bar and 400 bar, also exhibit ASE under ns-pulse excitation with similar in threshold values, as shown in Figures~\ref{ASEns}c-f. By characterizing different positions on each sample, Figure~\ref{ASEnsThres} shows minimal change in ASE thresholds, suggesting that the recrystallization pressure at 300 bar is essentially adequate to achieve high-quality and high-gain solution-based CsPbBr$_3$ thin film.
    
    {\color{black}\subsection{Comparison of ASE threshold of the reported CsPbBr$_3$ thin films}}
    
    \begin{table}[h]
    \centering
    \caption{\color{black}{Comparison of ASE thresholds reported for CsPbBr$_3$ thin films in the literature.}}
{\color{black}
\begin{tabular}{lcccc}
\hline
Reference & Film type & Thickness (nm)  & ASE threshold ($\mu$J cm$^{-2}$) \\
\hline
This work & Recrystallized thin film & $\sim115$  & $\sim200$ \\
~\cite{Wang2020b} & Polycrystalline thin film (doped) & $\sim210$ & 271--399 \\
~\cite{Wang2020b} & Polycrystalline thin film (pristine) & $\sim210$ & $\sim483$ \\
~\cite{APL_CsPbBr3_SCTF_2024} & Single-crystal thin film & --  & $\sim55.6$ \\
~\cite{Pourdavoud2019} & Recrystallized film & -- & $\sim12.5$ \\
~\cite{Huang2023} & Thermally evaporated CsPbBr$_3$:TPPO film (1-photon ASE) & $\sim120$ & $\sim6$ \\
~\cite{Huang2023} & Thermally evaporated CsPbBr$_3$:TPPO film (2-photon ASE) & $\sim120$ & $\sim210$ \\
~\cite{Tatarinov2023} & Recrystallized film + cavities & $\sim100$ & $\sim14$--16 \\
\hline
\end{tabular}
}
    \label{tab:ASE_comparison}
    \end{table}

    {\color{black} Table\ref{tab:ASE_comparison} lists the examples of CsPbBr$_3$ with their corresponding film thickness and ASE threshold. In this work, the lowest threshold is obtained for the recrystallized 0.4 M sample with a thickness of 115 nm, yielding an ASE threshold of $\sim$200 $\mu$J/cm$^2$, on par with those reported for solution-processed CsPbBr$_3$ thin films. For example, ASE thresholds of 271-399$\mu$J/cm$^2$ have been reported for $\sim$210 nm thick CsPbBr$_3$ films doped with ammonium cations, while pristine films exhibit thresholds around $\sim$483$\mu$J/cm$^2$.~\cite{OME_CsPbBr3_ASE_2020}. On the other hand, lower thresholds have been demonstrated in systems with improved crystalline quality or different optical geometries. For instance, centimeter-scale CsPbBr$_3$ single-crystal thin films exhibit ASE thresholds of $\sim$55.6$\mu$J/cm$^2$~\cite{APL_CsPbBr3_SCTF_2024}, reflecting reduced scattering losses and a lower density of non-radiative recombination centers compared with polycrystalline films. Even lower values have been reported in imprint-assisted recrystallized CsPbBr$_3$ layers, where ASE thresholds as low as $\sim$12.5$\mu$J/cm$^2$ were achieved due to enhanced film quality and reduced optical losses~\cite{Pourdavoud2019}. Additionally, Huang et al. reported thermally evaporated CsPbBr$_3$:TPPO films with an ASE threshold of 6~$\mu$J/cm$^2$ under one-photon excitation and 210~$\mu$J/cm$^2$ under two-photon excitation \cite{Huang2023}. These low thresholds arise from the combination of additive-assisted thermal co-evaporation and improved gain properties. Finally, very low lasing thresholds have been reported in high-quality films integrated into microcavity geometries such as microdisks, where optical feedback and reduced propagation losses further lower the threshold~\cite{Tatarinov2023}.}

    {\color{black} Several experimental factors contribute to the difference between these best-in-class values and the thresholds measured here. First, the active layer thickness in our recrystallized samples 60-115 nm is smaller than in many reports, where film thicknesses typically range from 150-250 nm. Thinner layers reduce both the pump absorption and the optical confinement factor of the guided mode, thereby decreasing the modal gain $g_{\mathrm{modal}}=\Gamma g_{\mathrm{mat}}$ and increasing the incident fluence required to reach the ASE condition. Second, the excitation beam in our experiments has a Gaussian spatial profile, so that the carrier density varies across the pumped region and only the central part of the beam reaches the threshold condition near the onset. As a consequence, the ASE emission originates from a smaller effective amplification region than the nominal pump area, which increases the apparent threshold when the fluence is defined using the total pulse energy and the nominal beam size. Similar effects associated with non-uniform excitation profiles have been discussed in the context of ASE and variable-stripe measurements in perovskite films~\cite{AlvaradoLeanos_AOM_2021}.}}

    \section{Phase transition of the recrystallized thin film of PbBr$_2$ precursor concentrations 0.23 M and 0.3 M} 
\begin{figure}[htb!]
		\begin{center}
			\includegraphics[width=16cm]{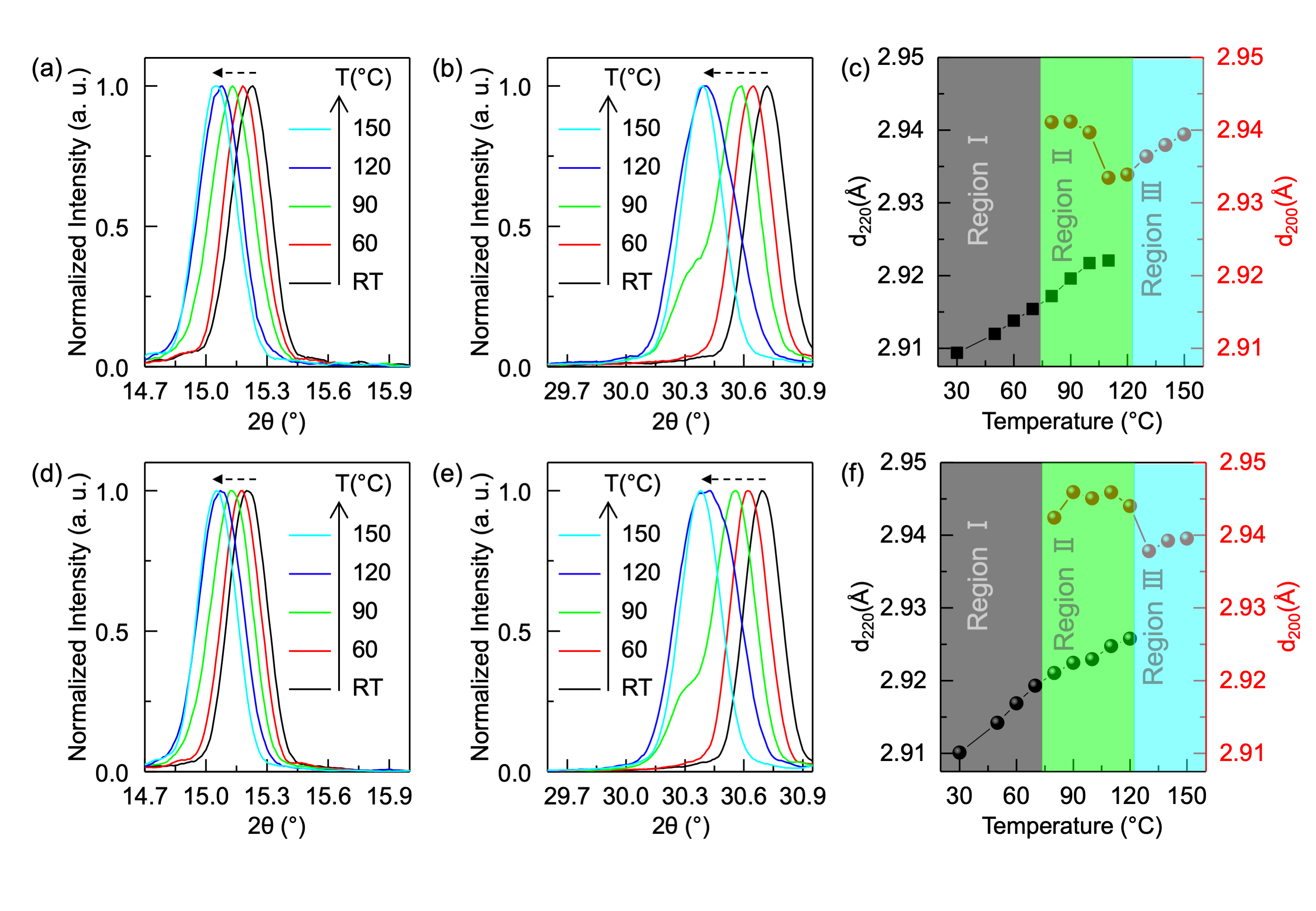}
		\end{center}
		\caption{Recrystallized perovskite layer of 0.23 M of PbBr$_2$  precursor concentration (a, b) XRD peaks at $2\theta = 15.21^\circ$ and $30.70^\circ$, respectively, change as a function of temperature (from room temperature to 150°C) and (c) the evolution of interplanar distances of the diffraction planes (220) and (200)  as a function of temperature. Recrystallized perovskite layer of 0.3 M of PbBr$_2$ precursor concentration (d, e) XRD peaks at $2\theta = 15.21^\circ$ and $30.70^\circ$, respectively, change as a function of temperature (from room temperature to 150°C) and (f) the evolution of interplanar distances of diffraction plane (220) and (200) as a function of temperature.}
		\label{PhaseTrans02303}
	\end{figure}

    Figures~\ref{PhaseTrans02303}a and b are the XRD primary peaks at $2\theta = 15.21^\circ$ and $2\theta = 30.69^\circ$ at different temperatures of the recrystallized PbBr$_2$ precursor concentration 0.23 M thin film, respectively. The results show the shift to smaller angles as the temperature increases for both peaks. Especially, the evolution corresponding to the primary peak at $2\theta = 30.69^\circ$ (Figure~\ref{PhaseTrans02303}b) shows a shoulder that appears at 90°C, indicating the phase changes from orthorhombic to tetragonal, and finally, cubic. This transition can be presented by the change of interplanar distances as a function of temperature, as shown in Figure~\ref{PhaseTrans02303}c. The complete transition to the cubic phase is achieved from around $120^\circ C$, which is 10°C lower than the case of PbBr$_2$ precursor concentration 0.4 M recrystallized thin films. Similar behavior is observed with PbBr$_2$ precursor concentration 0.3 M recrystallized thin film as demonstrated in Figures~\ref{PhaseTrans02303}d-f.
    
{\color{black} 

\end{document}